\documentclass[preprint]{elsarticle}

\usepackage{graphicx}
\usepackage{subfigure}
\usepackage{natbib}
\usepackage{amssymb}
\usepackage{amsmath}

\journal{New Astronomy}

\begin{document}

\begin{frontmatter}

\title{ANTARES -- A Numerical Tool for Astrophysical RESearch \\
       With applications to solar granulation }

\author[fmv]{H.J. Muthsam\corref{cor1}}
\ead{herbert.muthsam@univie.ac.at} 
\author[mpa,opm]{F. Kupka} 
\author[fmv]{B. L\"ow-Baselli}
\author[fmv]{C. Obertscheider}
\author[fmv]{M. Langer}
\author[fmv,ifav]{P. Lenz}

\cortext[cor1]{corresponding author}
\address[fmv]{Institute of Mathematics, University of Vienna, Nordbergstra{\ss}e 15,
                       A-1090 Vienna, Austria}
\address[mpa]{Max-Planck-Institute for Astrophysics, Karl-Schwarzschild Stra{\ss}e 1, 
                        D-85748 Garching, Germany}  
\address[ifav]{Institute for Astronomy, University of Vienna, T\"urkenschanzstra{\ss}e 17,
                      A-1180 Vienna, Austria}
\address[opm]{now at Observatoire de Paris, LESIA, CNRS UMR 8109, F-92195 Meudon, France}
  
%\titlerunning{ANTARES -- A Numerical Tool for Astrophysical RESearch}
%authorrunning{H.J. Muthsam et al.}

\begin{abstract}
We discuss the general design of the ANTARES code which is intended for simulations in stellar hydrodynamics with radiative transfer and realistic microphysics in 1D, 2D and 3D. We then compare the quality of various numerical methods. We have applied ANTARES in order to obtain high resolution simulations of solar granulation which we describe and analyze. In order to obtain high resolution, we apply grid refinement to a region predominantly occupied by an exploding granule. Strong, rapidly rotating vortex tubes of small diameter ($\sim 100\mbox{ km})$ generated by the downdrafts and ascending into the photosphere near the granule boundaries evolve, often entering the photosphere from below in an arclike fashion. They essentially contribute to the turbulent velocity field near the granule boundaries.
\end{abstract}


\end{frontmatter}

%----------------------------------------------------------------------------------------------
%---------------------------------------- INTRODUCTION ----------------------------------------
%----------------------------------------------------------------------------------------------
\section{Introduction}
\par
Modelling of hydrodynamic flows in the setting as appropriate for stellar physics has provided a large number of insights over the past decades. If we stick to the  vicinity of solar granulation studies only, we may mention work such as 
\cite{Nordlund82}, \cite{Stein98}, \cite{Wedemeyer04}, \cite{Robinson05}, \cite{Voegler05}.

In order to conduct research in the context of stellar convection and related areas we are developing a code for such simulations. In this paper, we describe the most central and most mature features of the continually expanding code, i.e. hydrodynamics with radiative transfer and realistic material properties for the stellar and of course solar case. We discuss the relative performance of various numerical methods. We exemplify the use and benefits granted by the program in discussing high resolution simulations of solar granulation.

ANTARES (A Numerical Tool for Astrophysical RESearch) is designed according to the  following general principles and aims:
\begin{itemize}
\item{{\bf general:} time dependent compressible hydrodynamics and extensions (such as MHD) in 1D, 2D and 3D; Fortran90 code with a modular structure;
}
\item{{\bf numerics:} various high resolution numerical schemes of conservative form implemented in order to evaluate their merits;
}
\item{{\bf radiative transfer:} short characteristics method (use of the diffusion approximation where appropriate); either grey or nongrey by the binning method based on state of the art opacities;
}
\item{{\bf microphysics:} realistic (or idealized);
}
\item{{\bf gridding:} logically rectangular; either rectangular or spherical coordinates; equidistant or (vertically) logarithmically spaced grid points; grid refinement in pre-assigned rectangular patches, even recursively;
} 
\item{{\bf parallelization:} via MPI (additional OpenMP nearing completion);
}
\item{{\bf portability:} is running on
	\begin{itemize}
	\item{AMD, IBM and Intel processors}
	\item{AIX, Linux and OSX operating systems.}
	\end{itemize}
}
\end{itemize}

The use of high resolution numerics, together with grid refinement, has of course the aim of achieving the best possible \emph{effective resolution}. Given that just this is the aim of so many efforts on the observational side, this seems to be a worthwhile endeavor also on part of numerical modelling. Regarding observations in the solar case, we mention, for example, the Swedish Vacuum Telescope or the Sunrise project \citep{Gandorfer04}.

We have described 2D high resolution simulations of solar granulation in a previous paper \citep{Muthsam07}. There the main result was the occurrence of strong acoustic pulses generated by the granular downdrafts and moving over considerable length, i.e., one, two or even more granule diameters. Furthermore, we observed strong vortex patches generated by the integranular downdrafts which remained, however, quite stably near the place where they had been generated and in any case well below the photosphere. -- We should be able to see such phenomena in the resolution we achieve in the grid refinement region in the 3D calculation which we discuss in the present paper. The question which we discuss concerns therefore the nature of 3D solar granulation at high resolution.

%---------------------------------------- RHD EQUATIONS ----------------------------------------
\section{The equations of radiation hydrodynamics (RHD)}

With the ANTARES code, the equations of radiation hydrodynamics are solved in one, two, or three dimensions. 

%---------------------------------------- HD ----------------------------------------
\subsection{The hydrodynamical equations}

The equations governing the dynamics of granulation or convection in the Sun or in stars (without magnetic activity for the present purpose), are, see e.g. \citet{Mihalas84}, the continuity equation
\begin{equation}
\partial_{t}\rho + \nabla \cdot \left(\rho {\bf u}\right) = 0,  \label{cont-eq}
\end{equation}
the momentum equation
\begin{equation}
\partial_{t}(\rho{\bf u}) + \nabla \cdot (\rho {\bf u}{\bf u} +p \underline{I}) = \rho{\bf g}  + \nabla \cdot \underline{\tau},  \label{mom-eq}
\end{equation}
and the total energy equation
\begin{equation}
\partial_{t}e + \nabla \cdot ({\bf u}(e+p)) = \rho({\bf g}\cdot {\bf u})  + \nabla \cdot ({\bf u} \cdot \underline{\tau}) + Q_{\rm{rad}}.  \label{energy-eq}
\end{equation}
All physical quantities are functions of space and time coordinates, ${\bf x}=(x,y,z)^\ast$ and $t$, i.e. $\rho = \rho({\bf x},t)$. The x-direction points in the vertical direction, y- and z-direction are the horizontal coordinates. The meaning of the most important physical variables used here and
in the following is summarized in Table~\ref{tlab}.
\begin{table}[ht]
\begin{center}
\caption{Meaning of the main variables} \label{tlab}
\begin{tabular}{ll p{5cm}}
\hline
$\rho$                    & gas density \\
${\bf u}=(u,v,w)^\ast$  & flow velocity \\
$\rho{\bf u}$           & momentum density\\
${\bf u}{\bf u}$        & dyadic product \\
$p$                       & gas pressure \\
$\underline{I}$        & identity matrix \\
${\bf g}=(g,0,0)^\ast$  & gravitational acceleration \\
$\underline{\tau}$   & viscous stress tensor for zero bulk viscosity \\
& $ \underline{\tau}_{ij}=\mu \left (\partial_{x_j}u_i + \partial_{x_i}u_j -\frac{2}{3} \delta_{ij} (\nabla \cdot {\bf u})  \right)$ \\
$\mu$                   & dynamic (molecular) viscosity \\
$e=e_{\rm int}+e_{\rm kin}$ & `total' energy density, i.e.\ the sum \\
                            & of internal and kinetic energy densities \\
$T$                      & temperature \\
$Q_{\rm{rad}}$       & radiative source term \\
$\chi_{\nu}$           & (specific) opacity at frequency $\nu$ \\
$\kappa$               & radiative conductivity \\
\hline
\end{tabular}
\end{center}
\end{table}

%---------------------------------------- RT ----------------------------------------
\subsection{The radiative transfer (RT) equation}

Any realistic simulation of solar surface (photosphere and upper convection zone) flows
must include the energy exchange between gas and radiation. This process
is described by the radiative heating rate $Q_{\rm{rad}}$ which is an additive term in 
the energy equation.
\par
In order to determine $Q_{\rm{rad}}$ the stationary limit of the radiative transfer equation
\begin{equation}
{\bf r} \cdot \nabla I_{\nu} = \rho \chi_{\nu} (S_{\nu} - I_{\nu}), \label{rte}
\end{equation}
see \citet{Mihalas78a}, is solved for all ray directions ${\bf r}$ and for all frequencies $\nu$ resulting the specific 
intensity $I_{\nu}({\bf r})$. The stationary limit represented by (\ref{rte}) assumes that 
the temporal variation of the flow is slow and thus its associated timescale is long in comparison 
with that one of photon transfer. Flow velocities are required to be non-relativistic 
($|\bf{u}| \ll {\rm c}$). Moreover, the radiative energy density and its variation as a function
of time are assumed to be small in comparison with the total energy density $e$ and its
temporal variation. For a general discussion of the validity of these equations see \citet{Mihalas84}. -- 
Since local thermodynamic equilibrium is assumed to hold and scattering is omitted, the source
function is the Planck function,  $S_{\nu}=B_{\nu}$. The (specific) opacity $\chi_{\nu}$ at
given thermodynamical conditions $(\rho ,T)$ is approximated such that it can be 
interpolated from precomputed tables (see Sect.~\ref{binning}). 
If the radiation field is known, the mean intensity is
\begin{equation}
J_{\nu}=\frac{1}{4\pi} \int I_{\nu}( {\bf r} )d\omega
\end{equation}
and the radiative energy flux is
\begin{equation}
F_{\nu}= \int I_{\nu}( {\bf r} ) {\bf r} d\omega
\end{equation}
(see for example \citet{Mihalas78b}). The radiative heating rate representing the difference between absorption and emission can be determined either from
\begin{equation}
Q_{\rm{rad}} = 4 \pi  \rho \int_{_{\nu}}\chi_{\nu}(J_{\nu}-S_{\nu}) d{\nu}
\end{equation}
or from the equivalent expression
\begin{equation}
Q_{\rm{rad}} = - \int_{_{\nu}} (\nabla \cdot  F_{\nu}) d{\nu}.
\end{equation}

%---------------------------------------- EOS ----------------------------------------
\subsection{The equation of state \label{section-eos}}

The conservations laws for radiation hydrodynamics (\ref{cont-eq})--(\ref{energy-eq}) are closed by the equation of state (EOS) which describes the relation between the thermodynamic quantities. The EOS used takes into account 
partial ionization which is present in the uppermost few Mm of the Sun.
\par
In the simulations presented here realistic microphysics is included by the up-do-date OPAL equation of 
state \citep{Iglesias96}. 
\par
For grey radiative transfer Rosseland opacities given by 
        Iglesias and Rogers, \citep{Iglesias96}, extended at low temperature
        by Alexander and Ferguson, \citep{Alexander94} and (in all more recent
        simulations) by Ferguson et al., \citep{Ferguson05}, are adopted.  For non-grey radiative transfer parts of the ATLAS 9 package \citep{Kurucz93} are used (i.e.\ opacity distribution functions and model atmospheres)
to determine bin-averaged opacities and source functions. 
\par
For largely arbitrary chemical compositions, i.e. values of $X$, $Y$ and $Z$, the equation of state is precomputed for a logarithmic mass density-temperature grid which
provides all required thermodynamical quantities for a given pair ($\rho$,$T$).

\par
Since the system of equations (\ref{cont-eq})--(\ref{energy-eq}) is solved in time
to obtain mass density and total energy density, the internal energy density (total minus kinetic 
energy density) and temperature must be calculated. The latter is interpolated from a precomputed grid which provides temperature as a function of logarithmic mass density
and internal energy density. Alternatively and more precisely, $T$ can be obtained by a Newton-Raphson procedure based on the table mentioned in the last paragraph.
\par
Bin-averaged quantities for non-grey radiative transfer are interpolated in a precomputed logarithmic mass density vs. pressure grid.

%----------------------------------------------------------------------------------------------
%---------------------------------------- ANTARES CODE ----------------------------------------
%----------------------------------------------------------------------------------------------
\section{The ANTARES code}

The ANTARES (Advanced Numerical Tool for Astrophysical RESearch) code as described here can perform compressible hydrodynamic simulations with full radiative transfer for the 1D, 2D, and
3D case on a rectangular grid, all comprised in one parallelized Fortran90 program. (Spherical coordinates with a logically rectangular grid are possible. We concentrate on the straight, equidistant grid here as it applies to the simulations which we analyze later on in that paper.) -- Various high-resolution numerical methods (all of the essentially non oscillatory, ENO, type, see \citet{Liu94},
\citet{Fedkiw98}, \citet{Liu98}) are implemented. The order of spatial and temporal discretization can be chosen arbitrarily to a considerable extent. The viscous terms can be discretized by a fourth-order accurate scheme or be replaced by artificial diffusivities \citet{Stein98}, \citet{Caunt01}. Optionally, we can also use the Smagorinsky subgrid scale model, \cite{Smagorinsky63}. -- Furthermore, the ANTARES code allows local grid-refinement in rectangular patches, even recursively. Data from the refinement are projected up to the coarser level.

%---------------------------------------- STRUCTURE ----------------------------------------
\subsection{Structure of the ANTARES code}

Given the physical state ${\bf q}^n$ at time step $n$ a Runge-Kutta step is performed to get the physical state ${\bf q}^{n+1}$ at time $n+1$. Each substep consists of the following parts:
\begin{itemize}
\item Approximation of the spatial derivatives of the hyperbolic part with ENO-type schemes (Sect.~\ref{section-eno}).
\item Discretization of the viscosity: fourth-order discretization of the viscous stress tensor or artificial diffusivities (Sect.~\ref{section-viscosity}).
\item Determination of the radiative heating rate (Sect.~\ref{section-numrt}).
\item Evaluation of all other source terms.
\item Update the values for $\rho$, $\rho{\bf u}$ and $e$ according to the temporal integration (Sect.~\ref{section-rk}).
\item Call the equation of state to get all other used thermodynamical quantities 
(Sect.~\ref{section-eos}).
\item Apply the boundary conditions for subsequent use (see Sect.~\ref{section-bc}).
\end{itemize}

%---------------------------------------- GRID ----------------------------------------
\subsection{Numerical grid}

The region of interest $D$ for the presented numerical simulation is a rectangular domain. The rectangular computational domain is equipped with a uniform numerical grid: $N_x$ gridpoints in x-direction, $N_y$ gridpoints in y-direction and $N_z$ gridpoints in z-direction with up to seven ghost cells at each boundary, depending on the order of the spatial discretization. The mesh sizes are $\Delta x$, $\Delta y$ and $\Delta z$. (In addition, the ANTARES code also offers the option for logarithmic grid spacing.)
\par
The vertical $x$-range covers $x \in [0,x_{\max}]$ ($x=0$ is at the top), consisting of $N_x$ intervals with length $\Delta x=x_{\max}/N_x$. Similarly, the horizontal ranges cover $y \in [0,y_{\max}]$ and $z \in [0,z_{\max}]$, consisting of $N_y$ resp. $N_z$ intervals with lengths $\Delta y=y_{\max}/N_y$ resp. $\Delta z=z_{\max}/N_z$. 
\par
The grid points $(x_i,y_j,z_k)$ are the centers of the cells $I_{i,j,k}$. For $i=1,\ldots N_x$, $j=1,\ldots N_y$ and $k=1,\ldots N_z$

\begin{equation*}
x_i = (i-\frac{1}{2}) \cdot \Delta x \mbox{, }\hspace{0.3cm}  
y_i = (j-\frac{1}{2}) \cdot \Delta y \mbox{, }\hspace{0.3cm}  
z_k = (k-\frac{1}{2}) \cdot \Delta z \hspace{0.3cm}\mbox{, }\hspace{0.3cm}
\end{equation*}
and
\begin{equation*}
 I_{i,j,k}=[x_{i-\frac{1}{2}}, x_{i+\frac{1}{2}}] \times [y_{j-\frac{1}{2}}, y_{j+\frac{1}{2}}] \times [z_{k-\frac{1}{2}}, z_{k+\frac{1}{2}}] 
\end{equation*}
where
\begin{equation*}
x_{i \pm \frac{1}{2}} = x_i \pm \frac{\Delta x}{2} \mbox{, }\hspace{0.3cm}
y_{j \pm \frac{1}{2}} = y_j \pm \frac{\Delta y}{2} \mbox{, }\hspace{0.3cm}
z_{k \pm \frac{1}{2}} = z_k \pm \frac{\Delta z}{2}\mbox{ .}  
\end{equation*}

%---------------------------------------- GRID REFINEMENT ----------------------------------------
\subsection{Grid refinement}

\begin{figure}[ht]
\begin{center}
\includegraphics[width=50mm]{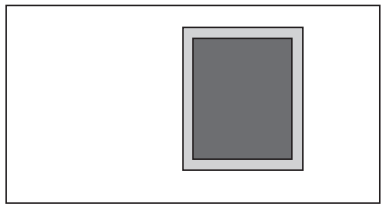}
\caption{Domains for a two-dimensional grid refinement.}
\label{gr-domain}
\end{center}
\end{figure}

For detailed studies, for example a single granule or the region near a downflowing plume, local grid refinement can be used. First, the region of interest for the refined area (dark grey region in 
Fig.~\ref{gr-domain}) and the (integer) grid refinement factor (for each direction) must be specified. The refined area is surrounded by ghost cells (light grey region in Fig.~\ref{gr-domain}) to allow regular interpolation and symmetric differentiation near the boundaries. Then, the values of the physical quantities at the coarse grid are interpolated to the fine grid and the ghost cells (light and dark grey region). This yields the initial state for the simulations. Furthermore, a time step for the time evolution at the fine grid $\Delta t_{GR}=\frac{\Delta t}{N}$ for an integer $N$ is calculated such that all numerical time step restrictions are fulfilled. Typically, $\Delta t$ is 
divided by the grid refinement factor in the vertical direction unless there are further constraints due to time step restrictions.
\par
Each step of a simulation with a grid refinement region consists of a Runge-Kutta time step on the whole coarse grid. Then, the initial data at the coarse grid are interpolated to the ghost cells surrounding the refined area and $N$ Runge-Kutta steps are performed on the fine grid. Before each Runge-Kutta step or intermediate step, which corresponds to a time $t_{\rm GR} \in [t,t+\Delta t]$, the ghost cells are assigned with linear interpolation values from the physical states ${\bf q}(t)$ and ${\bf q}(t+\Delta t)$ from the coarse grid. The values at the grid points of the coarse grid which belong to the refined area are overwritten by a volume-integrated approximation from the solution ${\bf q}_{GR}(t+\Delta t)$ at the fine grid after N Runge-Kutta steps for the
latter (back projection onto the coarser level). 

The back projection step causes a slight violation of conservation. To be more precise: in the interior of the refinement region the back projection is strictly conservative. However, at the boundaries of the refinement region the borders of the refined cells will not necessarily coincide with the borders of the coarse cells. There, a non-conservative sort of back projection is presently used. No unfavourable effects (such as noticeable mass loss) have been observed.
Otherwise, the numerics proper is of the conservation type so that mass, momentum and energy are conserved (except for flow of those quantities through the upper or lower boundary). Of course, conservation does also not apply to the $Q_{rad}$ term in Eq.~\ref{energy-eq} in those regions where the radiative transfer equation (and not the diffusion approximation) is used, since this term cannot be reasonably cast into divergence form in the present context.

%---------------------------------------- PARALLELIZATION ----------------------------------------
\subsection{Parallelization}

\begin{figure}[ht]
\begin{center}
\includegraphics[width=50mm]{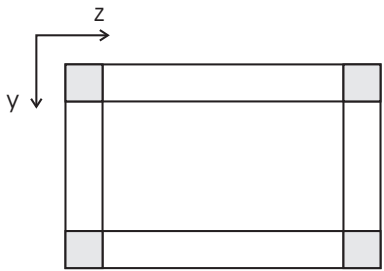}
\caption{Corners in the ($y \times z$)-plane.}
\label{corners}
\end{center}
\end{figure}

Parallelization is achieved with MPI according to the distributed memory concept. Each processor performs a simulation on a rectangular subdomain. The number of required processors is specified by the number of subdivisions in \mbox{x-}, \mbox{y-}, and z-direction 
($P_x,P_y,P_z$). The resulting subdomains are equipped with ghost cells in all directions. If an edge area of a subdomain is adjacent to another subdomain, the ghost cells are filled with the values from the neighboring subdomain. Each (intermediate) state is distributed to the adjacent subdomain. If an edge area represents a physical boundary then the physical boundary conditions are applied. Corner values (see Fig.~\ref{corners}) in any ($y \times z$)-plane (also in the ghost regions in x-direction) are interpolated, if required.
\par
\begin{figure}[ht]
\begin{center}
\includegraphics[width=80mm]{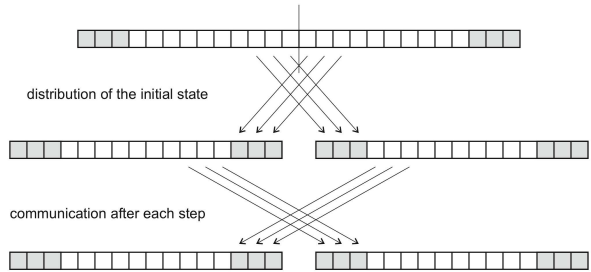}
\caption{Domain decomposition in two subdomains for the one-dimensional case.}
\label{domaindecomposition}
\end{center}
\end{figure}

The one-dimensional decomposition procedure is illustrated in Fig.~\ref{domaindecomposition}. The whole region (physical (white) and ghost (grey) cells) is split up. For each subdomain additional ghost cells are introduced. These ghost cells are filled with the values of the adjacent domain.
\par
The details of the parallelization of the non-local radiative transfer treatment are described in Sect.~\ref{nrt}.
\par
For the parallelization of a model with a grid refinement zone, the coarse and the fine grid are 
split to the processors according to the distributed memory concept. Hence, the computational
effort is uniformly distributed among all processors.

%---------------------------------------- PERFORMANCE ----------------------------------------
\subsection{Performance}

The ANTARES code has been run successfully on up to 256 processors in full
RHD simulations using MPI parallelization only. Replacing for test purposes the 
radiative heating rate $Q_{\rm rad}$ by the diffusion approximation the algorithm
scales almost perfectly as long as the subdomains on each processor are large
enough such that the ghost cells are a small part of the computational domain and
thus the data transfer between the nodes remains small.
\par
Since the radiative transfer is non-local, a full RHD simulation cannot scale perfectly. Our implementation of the radiative transfer (see Sect.~\ref{section-numrt}) for example yields a theoretical efficiency of 80\% (neglecting data transfer and omitting the transition to the diffusion approximation) for a 3D simulation with $\Delta x=\Delta y=\Delta z$ on
64 processors ($P_x=P_y=P_z=4$). In our RHD simulations the code is mostly run, 
however, with $P_x$ equal to $1$ or $2$, because the diffusion approximation can
be used for the computation of $Q_{\rm rad}$ for those layers of the model domain
which always remain optically thick. The optimal choice of $P_x$, $P_y$, and $P_z$
is thus a compromise that minimizes total computational costs, idling due to
different physical complexity and hence different equations solved within the 
computational domain (RHD equations vs.\ diffusion approximation), and idling due to 
data transfer between processors.

%---------------------------------------- STARTING MODEL ----------------------------------------
\subsection{Initial conditions}

Typically, we use a slightly perturbed static model atmosphere or envelope which we equip with a small seed velocity field or pressure perturbations etc.  when starting a model. For example, for 2D simulations of solar convection described in \citet{Muthsam07} and furthermore discussed below we start from plane parallel models of the solar 
atmosphere and an envelope due to \citet{JCD96}, with a small velocity field applied. For 3D simulations a converged 2D physical state is converted into a 3D one by putting copies of the two-dimensional model side-by-side. Again, velocities are 
perturbed slightly so as to introduce a truly 3D field and then the model is evolved in time.
Overall, this reduces the computational costs for relaxing 3D simulations to a statistically
quasi-stationary state by up to about 50\%.

%---------------------------------------- BC ----------------------------------------
\subsection{Boundary conditions \label{section-bc}}

All quantities are assumed to be periodic in both horizontal directions. 

%---------------------------------------- HD BC ----------------------------------------
\subsubsection{Hydrodynamical boundary conditions}

The solar simulations are currently performed with closed boundary conditions at the upper and 
lower boundary of the computational domain. These boundary conditions lead to unphysical 
reflections of waves and shocks. Thus, the results are not realistic in the upper photosphere. Presently, our interest 
focusses on lower layers, particularly around and below the superadiabatic layer, which 
essentially are unaffected \citep{Kupka08}.
\par
For closed boundaries at the top of the computational domain it is assumed that the vertical
fluxes of mass, horizontal momentum, and internal energy vanish. This is achieved by
setting the vertical velocity and the vertical gradients of mass density and internal energy
density to zero there:
\begin{equation}
u=0\mbox{,}\hspace{0.3cm}\partial_x\hspace{0.05cm}{\rho}=0\mbox{,}\hspace{0.3cm},
\partial_x\hspace{0.06cm}{e}=0\hspace{0.5cm}\mbox{ (at the top).}
\end{equation}
On the horizontal velocity components stress-free boundary conditions are applied:
\begin{equation}
 \partial_x\hspace{0.05cm}v  \equiv  0 \mbox{,}\hspace{0.3cm}
 \partial_x\hspace{0.05cm}w  \equiv  0 \hspace{0.5cm}\mbox{ (again at the top).}
 \end{equation}
\par
At the lower boundary the conditions for momentum read:
\begin{equation}
u \equiv 0\mbox{,}\hspace{0.3cm}\partial_x\hspace{0.05cm}v  \equiv  0 \mbox{,}\hspace{0.3cm}\partial_x\hspace{0.05cm}w  \equiv  0 \hspace{0.5cm}\mbox{ (bottom).}
\end{equation}
\par
If the bottom of the simulation is placed inside a convection zone, as is indeed always the case for solar surface convection, the implemented lower boundary conditions do not allow fluid motion through this lower boundary. Thus the convective and the kinetic energy flux densities are zero there. In order to have the proper energy being transported into the computational domain, we introduce a diffusive radiative flux $F_{\rm d}=\kappa^{*} \nabla T$. $\kappa^{*}$ is chosen such that the introduced diffusive radiative flux approximates the total energy flux at the bottom ($F_{\rm d} \rightarrow F_{\rm tot}$), and such that $\kappa^{*} \rightarrow \kappa$, the physical $\kappa$, three points inside the simulation domain ($F_{\rm d}=\kappa^{*} \nabla T$ converges to $F_{\rm rad}$). This procedure replaces the explicit boundary condition on $e$ at the bottom. 
\par
At the top of the domain three ghost cells are used to implement the boundary conditions. The physical state of the horizontal layer with $i=1$ is continued vertically to the ghost cells for mass, energy, $y$- and $z$-momentum densities. The ghost cell value for the $x$-momentum density is set to 0.
\par
At the bottom one ghost cell is used to implement the momentum boundary conditions. $y$- and $z$-momentum densities are continued vertically, the $x$-momentum density is set to zero. To adjust the radiative flux density the vertical temperature gradient at each point is constant in time and $\kappa$ is increased at three grid points to get the desired radiative flux density. 

No simple rule seems to exist which allows the boundary conditions to be implemented in a stable way. So, the exact number of ghost cells actually applied to the variables at the top or bottom has emerged from numerous experiments.
\par
Due to the implemented boundary conditions mass density and energy density are conserved.
\par
High order numerical methods may be in need for more ghost cells than we provide. If so, we reduce the order of the method near the boundaries in question. Alternatively, we resort to asymmetric stencils.

%---------------------------------------- RT BC ----------------------------------------
\subsubsection{Radiative boundary conditions} 

To solve the RTE in the computational domain boundary conditions for incoming radiation must be specified. 
\par
All quantities are assumed to be periodic in the horizontal directions. In the parallel mode, data transfer between the nodes is performed after each Runge-Kutta (sub-)step.  Regarding radiative transfer, we need the intensity values at the horizontal boundaries of each subdomain. These values are taken from the previous (sub-)step. This small time-lag does not noticeably affect the solution.  
\par 
At the top of the computational domain it is assumed that there is no incoming radiation, i.e. $\left. I_{\nu} ({\bf r}) \right|_{\rm top} =0$.
\par
At the bottom of the computation domain the diffusion approximation is valid, hence $\left. I_{\nu} ({\bf x},{\bf r}) \right|_{\rm bot} =B_{\nu}({\bf x})$. It is valid up to optical depths greater than about 10. In all applications we consider the optical depth of the lower boundary is orders of magnitude higher than that.
\par
To save computational time the lower boundary condition is applied at a fixed optical depth $\tau_{\rm DA} >10$ (calculated with Rosseland mean opacities) and in regions with optical depth greater than $\tau_{\rm DA}$ the radiative heating rate $Q_{\rm{rad}}$ is calculated according to the diffusion approximation $Q_{\rm{rad}}= \nabla \cdot (\kappa \nabla T)$.

%---------------------------------------- TIME STEP ----------------------------------------
\subsection{Time step restriction}

The maximal allowed time step is determined by the CFL-condition. In a single time step the flow is not allowed to transport information by more than one mesh width by both motion or sound waves. Thus, we get a time step restriction
\begin{equation}
\Delta t_{{\rm CFL}} \leq C_{{\rm Courant}} \cdot \frac{\min_i (\Delta x_i)^2}{\max(|{\bf u}|) + \max(c_{{\rm sound}})} 
\end{equation}
where $C_{{\rm Courant}}$ is the Courant number and $c_{{\rm sound}}$ is the sound speed. The $\max$ in the denominator is taken over the whole domain. $C_{{\rm Courant}}$ is set to $\frac{1}{4}$.
\par
The viscous stress tensor results in the time step restriction
\begin{equation}
\Delta t_{{\rm diffusive}} \leq C_{{\rm diffusive}} \frac{\min_i (\Delta x_i)^2}{ \frac{4}{3}\max \nu}.
\end{equation}
If artificial diffusivities (see Sect.~\ref{dovf}) are used, an additional diffusive time step
restriction is applied,
\begin{equation}  \label{eqhypdiff_restr}
\Delta t_{{\rm diffusive}} \leq C_{{\rm diffusive}} \frac{\min_i (\Delta x_i)^2}{\max \nu}.
\end{equation}
$C_{{\rm diffusive}}$ is also $\frac{1}{4}$. Again the $\max$ in the denominator is taken over the whole domain. The argument of this $\max$ are all shock stabilizing and hyperdiffusive viscosity coefficients. If subgrid-scale viscosities were used, time step restrictions would occur in a similar
manner and thus also accounted for as in equation \ref{eqhypdiff_restr}.

%------------------------------------------------------------------------------------------
%---------------------------------------- NUMERICS ----------------------------------------
%------------------------------------------------------------------------------------------
\section{Numerical methods}

The conservation equations (\ref{cont-eq})--(\ref{energy-eq}) are treated by the finite
volume method. The values of the fluxes are interpolated to the cell boundaries and
then the divergence for each cell is computed. The radiative heating rate is the most 
expensive part of the whole simulation and is determined by the short characteristics method for the RTE. All other source terms are simply evaluated at the
cell centers. Furthermore, the interpolation tool used and the temporal discretization
are described.

%---------------------------------------- (C)ENO ----------------------------------------
\subsection{Numerical schemes for conservation laws \label{section-eno}}

The spatial derivatives of the hyperbolic part of the system of conservation laws (\ref{cont-eq}), (\ref{mom-eq}), (\ref{energy-eq}),
\begin{equation}
\partial_t\hspace{0.03cm}{\bf q} +\partial_x\hspace{0.02cm}f({\bf q}) +\partial_y\hspace{0.02cm} g({\bf q})
 +\partial_z\hspace{0.03cm} h({\bf q}) =0, \label{conssys2}
\end{equation}
where
\begin{eqnarray} 
{\bf u} &=& (u,v,w)^\ast \nonumber \\
{\bf q} &=& (\rho,\rho u,\rho v,\rho w,e)^\ast \nonumber \\
f({\bf q}) &=& (\rho u, \rho u^2+p, \rho uv, \rho uw, u(e+p))^\ast \nonumber \\
g({\bf q}) &=& (\rho v, \rho vu, \rho v^2+p, \rho vw, v(e+p))^\ast \nonumber \\
h({\bf q}) &=& (\rho w, \rho wu, \rho wv, \rho w^2+p, w(e+p))^\ast,\nonumber 
\end{eqnarray}
are calculated for each direction separately. Approximations $\hat{f}_{i \pm \frac{1}{2},j,k}$, $\hat{g}_{i,j \pm \frac{1}{2},k}$ and $\hat{h}_{i,j,k \pm \frac{1}{2}}$ at the cell boundaries are evaluated in order to get approximations for 
\begin{eqnarray}  \partial_x\hspace{0.02cm}f({\bf q}) &\approx& \frac{1}{\Delta x}(\hat{f}_{i + \frac{1}{2},j,k}-\hat{f}_{i - \frac{1}{2},j,k}) \nonumber \\
 \partial_y\hspace{0.02cm} g({\bf q})&\approx& \frac{1}{\Delta y}(\hat{g}_{i,j + \frac{1}{2},k}-\hat{g}_{i,j - \frac{1}{2},k}) \nonumber \\ 
\partial_z\hspace{0.03cm} h({\bf q}) &\approx& \frac{1}{\Delta z}(\hat{h}_{i,j,k + \frac{1}{2}}-\hat{h}_{i,j,k - \frac{1}{2}}) \nonumber 
\end{eqnarray}
at each point $x_{i,j,k}$.
\par
The values for the numerical flux function at the cell boundaries are calculated for each component using one of the following high-resolution one-dimensional methods:
\begin{itemize}
\item Fourth-order convex non-oscillatory scheme (CNO-4), see \cite{Liu98}
\item Fifth-order weighted essentially non-oscillatory (ENO) scheme (WENO-5), see \citet{Harten78}, \citet{Shu88}, optionally in combination with Marquina flux splitting \citet{Donat96} 
\item Third-order essentially non-oscillatory scheme (ENO-3), cf.  \citet{Fedkiw98}, \citet{Donat96}
\end{itemize}

\subsubsection{Essentially non-oscillatory (ENO) and weighted essentially non-oscillatory (WENO) schemes}

Numerical methods for systems of nonlinear conservation laws must capture steep gradients (shocks and contact discontinuities) that may spontaneously develop and  persist in the solution. Classical numerical schemes either produce oscillations near steep gradients or smear out these gradients as well as fine details, see e.g. \cite{Leveque92}.
\par
The ENO type methods use adaptive stencils to avoid oscillations in the computed solution. Typically, when interpolating from cell centers to cell boundaries, around the cell boundary point to which one wants to interpolate the flux functions, the stencil consisting of neighbouring cell centers is selected or given maximum weight which leads to the smoothest interpoland for the flux functions using the stencil points. In addition, upwind considerations are applied.
\par
For systems of conservation laws this interpolation must be done in the local characteristic field since these quantities properly propagate in unique directions whereas the changes in density (say) will come about by signals travalling along the $0$ and the $+$ and $-$ characteristics. They arrive, in subsonic flow, from both sides and no meaningful upwind philosophy can therefore be devised in the original dependent variables.
\par
Speaking about the $x-$direction term in Eq.~\ref{conssys2}, the eigenvalues and eigenvectors of the Jacobian of flux the function $f'({\bf q})$ at point $x_{i+\frac{1}{2},j,k}$ (sic!) are determined. Then the flux function is transformed for sufficiently many neighboring points $x_{i-r:i+s,j,k}$ (again sic!) ($r+s+1=k$ is the order of the ENO approximation) to the local characteristic variables. Then, for each characteristic variable, the ENO algorithm is performed in order to obtain the cell boundary values at $x_{i+\frac{1}{2},j,k}$ for each component. The resulting cell boundary values are transformed back. This procedure is repeated similarly for the other directions.
\par
In the ANTARES code a fifth-order accurate weighted ENO scheme and a third-order accurate ENO scheme are implemented. For both methods optional use of Marquina's flux splitting and 
entropy fix, see \citep{Donat96}, is possible.

\subsubsection{Convex ENO (CNO) schemes}

Convex ENO schemes as described in \citet{Liu98} avoid the need to transform into the characteristic system. They are therefore simpler to implement and faster than ordinary ENO methods.  They achieve this by splitting the flux function $f$ into two components, $f=\frac{1}{2}(f^++f^-)$. The constitutents $f^\pm$ are designed in such a way that, for $f^+$ all characteristics run from the left to the right, and conversely for $f^-$. Thus, for each of them, there is a unique upwind direction and no need to go into the characteristic system in order to separate the two possible upwind directions. 
The convex ENO scheme uses, in addition, a reference flux, typically a modified second-order local Lax-Friedrichs flux, which reduces to first-order near discontinuities, and chooses the convex combination of the interpolation values of the different candidates stencils which is nearest to the reference flux. By the choice of a reference flux depending smoothly on the parameters, albeit of low order, Lax-Friedrichs in our case, and by this procedure smoothness of the resulting flux is achieved in addition to its high order due to the use of high order polynomials for interpolation.
\par
More specifically, we consider now the evolution equation for one of our basic dependent variables and describe the procedure obtaining cell boundary fluxes in direction $x$ (index $i$). For the various approximations (due to various choice of the stencils) at the cell boundaries for a component of the two flux functions for one component of our conservative system of equations 
\begin{equation}
f^+_{i+\frac{1}{2},j,k}(q)=\frac{1}{2} (f(q)+\alpha_{i+\frac{1}{2},j,k} q)
\end{equation}
and
\begin{equation}
f^-_{i+\frac{1}{2},j,k}(q)=\frac{1}{2} (f(q)-\alpha_{i+\frac{1}{2},j,k} q)
\end{equation}
where, in principle,  
\begin{equation}
\label{eq_cno_alpha}
\alpha_{i+\frac{1}{2},j,k}=\max_{ \min(q_{i,j,k},q_{i+1,j,k}) \leq q \leq \max(q_{i,j,k},q_{i+1,j,k}) } |f'(q)| 
\end{equation}
given by the ENO algorithm in $x$-direction, the CNO algorithm in the ANTARES code chooses for each sign $\pm$ the flux which is nearest to the Lax-Friedrichs reference flux, if all candidate fluxes are all larger or all  smaller than the reference flux. Otherwise, the reference flux is taken. 
\par
The choice of $\alpha_{i+\frac{1}{2},j,k}$ as in equation \ref{eq_cno_alpha} guarantees that, for each of the auxiliary fluxes $f^\pm$, the characteristics point all into one direction. In practices, $\alpha$ can be and is taken somewhat larger than the value just quoted resulting in greater stability at the expense of more smearing. 

\subsubsection{Test problem\label{testproblem}}

\begin{figure}[ht]
\begin{center}
\includegraphics[width=55mm,angle=270]{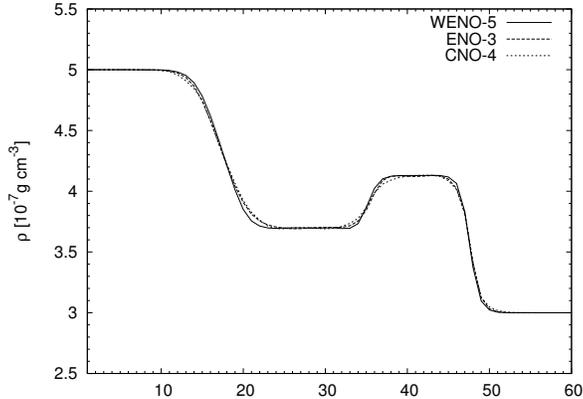}
\caption{Comparison of the numerical schemes.}
\label{num-dg}
\end{center}
\end{figure}

To show the main difference between these three schemes the one-dimensional version of the system (\ref{conssys2}) with the initial conditions
\begin{equation}
{\bf q}_l = \left(
\begin{array}{c} \rho _l  \\
u_l \\
T_l \end{array} \right) = 
\left( \begin{array}{c} 5.0\cdot 10^{-7}  \\
0  \\
11000 \end{array} \right) \; {\rm for} \; x \leq \frac{x_{\max}}{2} 
\end{equation}
\begin{equation}
{\bf q}_r = \left(
\begin{array}{c} \rho _r  \\
u_r \\
T_r \end{array} \right) = 
\left( \begin{array}{c} 3.0\cdot 10^{-7}  \\
0  \\
9000 \end{array} \right) \; {\rm for} \; x > \frac{x_{\max}}{2} 
\end{equation}
at 60 spatial grid points is evolved 100 second-order Runge-Kutta steps with $\Delta t=$0.7s. The resulting state is very similar for all this three high-resolution methods (see Fig.~\ref{num-dg}).
\par
\begin{figure}[ht]
\begin{center}
\includegraphics[width=55mm, angle=270]{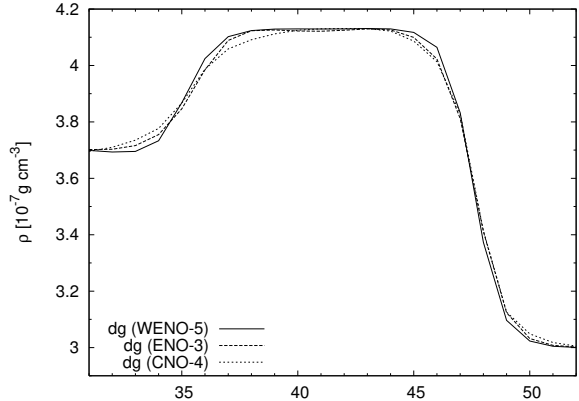}
\caption{Comparison of the numerical schemes for a region with steep gradients.}
\label{num-dg-detail1}
\end{center}
\end{figure}

A closer look at regions with steep gradients (see Fig.~\ref{num-dg-detail1}) shows some differences between the three methods. The fifth-order WENO scheme produces the steepest gradients. The CNO scheme is the most smoothing one. The third-order ENO solution lies between the two others in regions of steep gradients. The difference in the solution between the two different ENO-type schemes is due to the order of accuracy.

%---------------------------------------- VISCOSITY ----------------------------------------
\subsection{Viscous hydrodynamical fluxes \label{section-viscosity}}

In addition to the numerical diffusion of the ENO/CNO-schemes for the conservation laws the physical viscosity can be included by a fourth-order discretization of the viscous stress tensor or it can be replaced by artificial diffusivities.

\subsubsection{\label{dovf}Discretization of viscous fluxes}

The derivatives in the viscous stress tensor $\underline{\tau}$ are calculated by the fourth-order scheme
\begin{equation}
\left( \frac{\partial u}{\partial x} \right)_i=\frac{1}{12 \; \Delta x} \left( u_{i-2} -8 u_{i-1} +8u_{i+1} -u_{i+2} \right).   \nonumber
\end{equation}
Then, the values for $\underline{\tau}$ and $({\bf u} \cdot \underline{\tau})$ at the cell centers in the momentum and energy equations are interpolated to the cell boundaries by the following scheme:
\begin{equation}
\hat{f}_{i+\frac{1}{2}}=\frac{1}{16} \left( -f_{i-1} +9 f_{i} +9f_{i+1} -f_{i+2} \right)   \nonumber
\end{equation}
and the equivalent is done for $\hat{f}_{i-\frac{1}{2}}$
to get the contribution by $\nabla \cdot \underline{\tau}$ and $\nabla \cdot ({\bf u} \cdot \underline{\tau})$ through 
\begin{equation}
\frac{1}{\Delta x} \left( \hat{f}_{i+\frac{1}{2}}-\hat{f}_{i-\frac{1}{2}} \right) . \nonumber
\end{equation}

For simulations with this implementation of the viscous stress tensor we can choose a Prandtl number. Note that the Prandtl number typical for the solar atmosphere, see \cite{Komm91},  $\rm{Pr}=\frac{c_p \mu}{\kappa}$ varies from $1.4 \cdot 10^{-8}$ at the 
bottom to $1.0 \cdot 10^{-10}$ at the top of the computational domain. For our actual choice see the discussion at the end of subsection \ref{sec:numerics-discuss}. -- We note that the
radiative Prandtl number defined via $\mu_{\rm rad} = \rho \nu_{\rm rad} = 3 T \kappa / (4 c^2)$, cf. \cite{Kippen94}, is generally larger in the layers near the photosphere than the 
molecular viscosity; the latter has e.g. been calculated in \cite{Edmonds57}. For our simulations 
of solar granulation $\rm{Pr}_{\rm rad}$ ranges from a few times $10^{-10}$ to about 
$2 \cdot 10^{-8}$. As $\rm{Pr}_{\rm rad}$  can easily be calculated from equation of state related
quantities and opacities which are already required to calculate the radiative energy transfer, 
only the radiative viscosity is used in the simulations presented here. Hence, our $\rm{Pr}$
should more accurately be called $\rm{Pr}_{\rm rad}$, since conductive contributions to
$\kappa$ are negligible for the solar surface layers. It is important to not confuse this
definition of $\rm{Pr}$ with a purely molecular Prandtl number that compares 
molecular viscosity with heat conductivity: in most regions inside stars, typically as long as
the equation of state describes a perfect, non-degenerate gas, the heat conductivity is orders
of magnitudes smaller than the radiative one. Indeed, in stars the Prandtl number
defined through $\rm{Pr}_{\rm cond} = \frac{c_p \mu}{\kappa_{\rm cond}}$ is much larger than
$\rm{Pr}_{\rm rad}$ and much closer to that one of air. Physically, $\rm{Pr}_{\rm cond}$ is 
irrelevant, because either radiation or convection or both outperform conduction in heat transport 
inside the Sun by orders of magnitudes.
\par
Finally, we note that in those cases where we only use a physical viscosity term  instead of one that
is enhanced by an artificial diffusivity model, the numerical viscosity of the method for the 
hyperbolic part must be able to stabilize the simulation.

\subsubsection{Artificial diffusivities}

Artificial diffusivities as described in \cite{Stein98} and \cite{Caunt01} remove short-wavelength noise without damping longer wavelengths and diffuse strong discontinuities in order to stabilize the numerical code.
\par
The viscous stress tensor 
\begin{equation}
\underline{\tau}_{kl}=\mu \left( \partial_{x_l}u_k + \partial_{x_k}u_l -\frac{2}{3} \delta_{kl} (\nabla \cdot {\bf u})  \right) 
\end{equation}
is replaced by artificial equivalents of the form
\begin{equation}
\underline{\tau}_{kl}= \frac{1}{2} \rho \left( \nu_k({\bf{u}})\partial_{x_l}u_k  + \nu_l({\bf{u}})\partial_{x_k}u_l \right)
\end{equation}
where ${\rm u}=(u_1,u_2,u_3)^\ast$ and $k,l=1,2,3$.
\par
The coefficients $\nu_k$ for direction $k$ consist of two parts, a shock resolving $\nu_k^{\rm shk}$ and a hyperdiffusive $\nu_k^{\rm hyp}$:
\begin{equation}
\nu_k({\bf{u}})  =\nu_k^{\rm shk}({\bf{u}})+\nu_k^{\rm hyp}(u_k).
\end{equation}
The shock resolving part in direction $k$ is defined by
\begin{equation}
\nu_k^{\rm shk}({\bf{u}})= \left\{\begin{array}{ll} C_{\rm shk} \Delta x_k^2 |\nabla \cdot {\bf u} | & \nabla \cdot {\bf u} < 0 \\
0 & \nabla \cdot {\bf u} \geq 0 \end{array} \right.
\end{equation}
and acts therefore in regions undergoing compression only, i.e. where $\nabla \cdot {\bf u} < 0$.
The hyperdiffusive part in direction $k$ is defined by
\begin{equation}
\nu_k^{\rm hyp}(f)= c_{\rm hyp} \Delta x_i c_{\rm tot} \frac{\max_3 | \Delta_k^3 {f} |}{\max_3 | \Delta_k^1 {f} |}.
\end{equation}
Thereby we have 
\begin{equation}
c_{\rm tot} = | {\bf u}| +c_{\rm sound}; \nonumber
\end{equation}
$\Delta_k^1 f$ denotes the first difference of a scalar quantity $f$ ($u_k$ in the actual application) in direction $k$; $\max_3$ indicates that the maximum is taken over three adjacent intervals. $\Delta_k^3 f$ designates a similar third difference.
 For all simulations presented here $c_{\rm shk}=1$ and $c_{\rm hyp}=0.05$. These values make reasonable contact with those used in \cite{Caunt01}, namely $c_{\rm shk}=2$ and $c_{\rm hyp}=0.05$, and in \cite{Voegler05}, $c_{\rm shk}=1$ and $c_{\rm hyp}=0.03$.
 
 In addition, we have implemented the Smagorinsky subgrid model, see \citet{Smagorinsky63}, for optional use.

%---------------------------------------- INTERPOLATION ----------------------------------------
\subsection{\label{interpolation}Interpolation tools}

We need two interpolation schemes: one for use in the equation of state tables (in order to get smooth derivatives) and in connection with grid refinement, and an other one, being applied in the radiative transfer solver (in order to get positive opacities etc. even in the presence of huge gradients), which we describe in turn.

\begin{figure}[ht]
\begin{center}
\includegraphics[width=40mm]{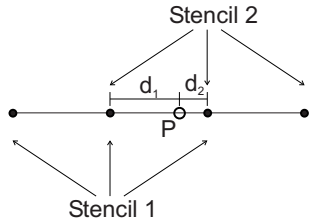}
\caption{Stencils for one-dimensional interpolation.}
\label{2d-int-stencil}
\end{center}
\end{figure}

\begin{figure}[ht]
\begin{center}
\includegraphics[width=40mm]{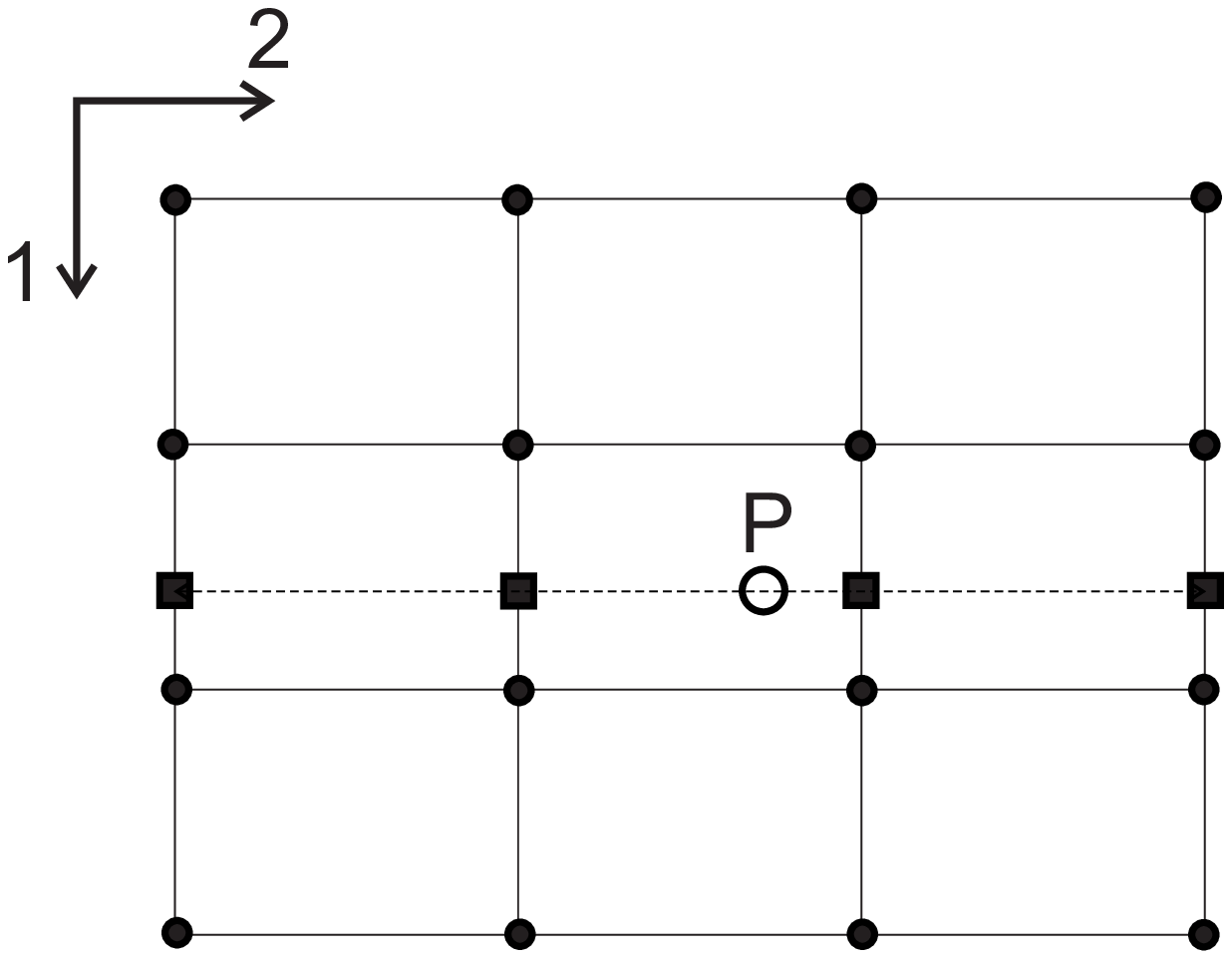}
\caption{Two-dimensional interpolation procedure.}
\label{2d-interpolation}
\end{center}
\end{figure}

For \emph{EOS and grid refinement interpolation} we use local splines \cite{Nendwich04}.  
Starting with four points, two polynomials through both of the three adjacent points are used to get two interpolation values at point $P$. These two results are combined by a weighted sum. The weight for polynomial 1 with stencil 1 is $\frac{d_2}{d_1+d_2}$ and for polynomial 2 with stencil 2 is $\frac{d_1}{d_1+d_2}$ (see Fig.~\ref{2d-int-stencil}).
%\par
In the two-dimensional case 4$\times$4 grid points are used to get the interpolation values. First, four one-dimen\-sional interpolations in direction $1$ and then one one-dimen\-sional interpolation in direction $2$ are performed to get the value at point $P$. The successive interpolation procedure is illustrated in Fig.~\ref{2d-interpolation}.
%\par
For three-dimensional interpolations 4$\times$4$\times$4 grid points are used and the procedure described above is performed starting with 16 interpolations in direction $1$, followed by 4 interpolations in direction $2$ and then one interpolation in direction $3$. This method yields
smooth (continuous) first derivatives everywher (cf. \cite{Nendwich04} for details on an efficient implementation also 
applicable to vector fields). 
\par
In the \emph{radiative transfer solver} a method for \emph{monotonic interpolation} in one dimension proposed by Steffen \cite{Steffen90} is applied. The method gives exact results if the data points correspond to a second-order polynomial. The application of this methods assures that the values for density, opacity, intensity, and the radiative source function always remain positive.

%---------------------------------------- NUMERICAL RT ----------------------------------------
\subsection{\label{nrt}Numerical radiative transfer \label{section-numrt}}

Instead of solving the three-dimensional RTE (\ref{rte}) the one-dimensional RTE
\begin{equation}
\partial_{\tau_{\nu}}I_{\nu} = S_{\nu} - I_{\nu}
\end{equation}
is solved along several ray directions, following the short characteristic method proposed by Mihalas \cite{Mihalas78b} and  Kunasz and Auer \cite{Kunasz88}. $d\tau_{\nu} = \chi_{\nu} \rho dx$ is the optical depth of a path element at frequency $\nu$ along a ray direction. 
\par
In the following the subscript $\nu$ which indicates the frequency dependence of $I$, $J$, $S$ 
and $\tau$ is dropped. Thus, the described numerical schemes treats grey radiative transfer. In Sect.~\ref{binning} frequency dependent RT is discussed.

\subsubsection{Short characteristic method} 

\begin{figure} [ht]
\begin{center}
  \includegraphics [width=5cm] {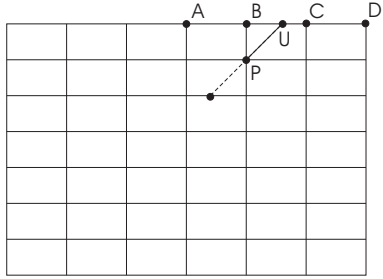}
\end{center} 
\caption{Short characteristic method, step 1 for ray directions entering at the top.}
\label{sc1}
\end{figure}

\begin{figure} [ht]
\begin{center}
   \includegraphics [width=5cm] {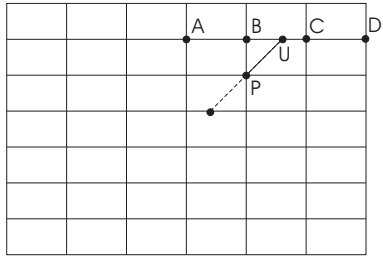}
\end{center} 
\caption{Short characteristic method, step 2 for ray directions entering at the top.}
\label{sc2}
\end{figure}

For incoming radiation the values for $I(\tau(U))$ along ray direction ${\bf r}$ are known. The optical depth $\tau(U)$ there is set to zero. To determine $I(\tau(P))$ along this ray direction ${\bf r}$ (see Fig.~\ref{sc1}) one interpolates the values of the required physical quantities to the point $U$ using 16 points in 3D and 4 points in 
2D, respectively. Then, one evaluates the equation
\begin{eqnarray}
I(\tau(P))&=& I(\tau(U)) {\rm e}^{\tau(U)-\tau(P)} + \nonumber \\
&& \int_{\tau(U)}^{\tau(P)} S(\tau') {\rm e}^{\tau'-\tau(P)}d\tau'
    \label{sc-formula}
\end{eqnarray}
numerically to get $I(\tau(P))$. This procedure is repeated recursively, since after step 1 the intensities for the first layer are determined (see Fig.~\ref{sc1} and~\ref{sc2}).
\par

\subsubsection{Numerical integration}

To perform the numerical integration on the right hand side of (\ref{sc-formula}) one has to determine $\tau(P)$ and $\tau(P+\overrightarrow{UP})$, where $\tau(U)$ is set to zero. The approximations are calculated by twice evaluating $\int \rho \chi dx$ using the trapezoidal rule.
\par
The integration on the right hand side of Eq. \ref{sc-formula} is performed by a quadrature rule proposed in \cite{Olson87}. 

\subsubsection{Angular integration}

\begin{table}[ht]
\begin{center}
\caption{Support points $x_i$ and the weights $a_i$ for the Gaussian quadrature rule for $N=10$ (see \cite{Lowan42})}
\label{20rays}
\begin{tabular}{cccc}
n=10 & i & $x_i$ & $a_i$  \\
\hline
& 1 & -0.974 & 0.067  \\
& 2 & -0.865 & 0.150  \\
& 3 & -0.680 & 0.219  \\
& 4 & -0.433 & 0.270  \\
& 5 & -0.149 & 0.296  \\
& 6 & 0.149 & 0.296  \\
& 7 & 0.433 & 0.270  \\
& 8 & 0.680 & 0.219  \\
& 9 & 0.865 & 0.150  \\
& 10 & 0.974 & 0.067  \\
\hline
\end{tabular}
\end{center}
\end{table}

For every grid point $x_{i,j,k}$ the RTE is solved along $N_{\rm rays}$ ray directions. To get the mean intensity $J$ an angular integration is performed.
\par
In the two-dimensional case the RTE is solved along 20 ray directions ${\bf r}_i=(r_{i1},r_{i2},r_{i3}=0)$. The $r_{i1}$ are set equal to the support points $x_i$ of the Gaussian quadrature rule for $N=10$ according to \cite{Lowan42}, the $r_{i2}$ are determined such that $r_{i1}^2+r_{i2}^2=1$ (for every $r_{i1}$ there are two solutions 
for $r_{i1}$, one with positive, one with negative sign). For the angular integration the corresponding weights $a_i$ are multiplied by $\pi$ since 
\begin{equation}
2\cdot\sum_{i=1}^{10} a_i =4\pi
\end{equation}
is required. The value of $4\pi$ instead of the possibly expected $2\pi$ stems from the fact what we consider to be the proper transport of radiation in 2D keeping the 3D case in mind as the ultimate goal. Namely, at a fixed location we do not only consider one ray of light (corresponding to a, say, positive horizontal direction and a certain zenith distance) within the computational plane but rather the half cone generated by such rays, which are allowed to move out of the plane of computation, pointing into positive horizontal direction and the same zenith distance). -- The values for $x_i$ and $a_i$ are shown in Table~\ref{20rays}.

In addition to this standard integration, ANTARES also offers the possibility to use 12 rays in 2D and to use the weights $a_i=2/({\rm N\_rays}/2)$.
\par
In the three-dimensional case the RTE is solved along 24 ray directions. The ray directions are chosen according to the angular quadrature formulae of type A of \cite{Carlson63}. The directions in each octant are arranged in a triangular pattern and the quadrature is invariant under rotations over multiples of $\pi/2$ around any coordinate axis. A summary of the construction procedure is given in \citet{Bruls99}. For the three-dimensional simulations the A4 quadrature set with three directions per octant is used. 
\par
For this choice each ray has got the weight $\omega_i=\frac{1}{24}$. The ray directions in the first octant are: \begin{align}
{\bf r}^1=\left(  \sqrt{7/9} , 1/3 , 1/3  \right)^\ast \nonumber \\
{\bf r}^2=\left(  1/3 , \sqrt{7/9} , 1/3  \right)^\ast \nonumber \\
{\bf r}^3=\left(  1/3 , 1/3 , \sqrt{7/9}  \right)^\ast \nonumber
\end{align}
The ray directions in the other octants are obtained applying three-dimensional rotations into the other octants.
\par
With the quadrature weights $\omega_i$ and the index $i$ running over the set of directions,
the mean intensity is
\begin{equation}
J = \sum_i \omega_i I({\bf r}^i).
\end{equation}

\subsubsection{Radiative heating rate}

With $J$ now calculated at each grid point the radiative heating rate can be obtained from
\begin{equation}
Q_{\rm rad} = 4\pi \rho \chi (J-S).
\end{equation}

\subsubsection{Implementation}

Radiative transfer is a non-local phenomenon. Simulations with $N_{\rm p}$ processors 
and MPI communication are carried out by the domain decomposition approach. Since 
the intensity is calculated with the short characteristic method along $N_{\rm rays}$ rays which start at a specific boundary of the computational domain, each processor has to wait until the information at the specific adjacent boundary is available.

% chapter 4.4.5
For simulations with numerous processors one has to ensure that as many processors as possible are busy. Our short characteristic algorithm classifies the $N_{\rm rays}$ rays, depending on the direction in which they pass the computational domain. 
\par
Considering the short characteristic algorithm, the rays are numbered such that the $N_{\rm rays,x}/2$ rays entering at the top of the cell are followed by the $N_{\rm rays,x}/2$ rays which enter at the bottom, followed by the $N_{\rm rays,y}/2$ rays entering at the lower y-boundary and by the $N_{\rm rays,y}/2$ rays entering at the upper y-boundary, followed by the $N_{\rm rays,z}/2$ rays entering at the lower z-boundary and by the $N_{\rm rays,z}/2$ rays entering at the upper z-boundary.
\par
For an equidistant three-dimensional grid the four ray directions entering at the top of the cell are called ${\bf r}^1, \ldots {\bf r}^4$. The four ray directions entering at the bottom of the cell are called ${\bf r}^5\ldots {\bf r}^8$. Of course, for the remaining four faces of the computational domain there exist similar directions ${\bf r}^9, \ldots {\bf r}^{24}$. 
\par
\begin{figure}[ht]
\begin{center}
\includegraphics[width=75mm]{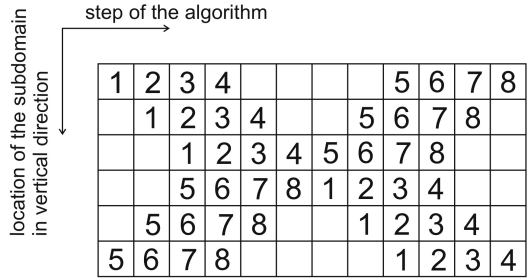}
\caption{Illustration of the implementation of the radiative transfer solver. Rays with number 1, 2, 3, 4 enter at the top of the cell, rays with number 5, 6, 7, 8 enter at the bottom.}
\label{rt-impl}
\end{center}
\end{figure}

Figure \ref{rt-impl} illustrates the implementation for the ray directions entering at the top and the bottom of the cell. The vertical axis is devided in $P_x=6$ parts which represent the 6 subdomains in $x$-direction. 
\par
The processors representing the upper half of the computational domain first calculate the intensities for the rays entering at the top of the cell (${\bf r}^1, \ldots {\bf r}^4$) in the following way: In the first step the processors representing the first layer from the top calculate the intensity for ${\bf r}^1$. In the second step these processors calculate the intensity for ${\bf r}^2$ and the processors representing the second layer from the top calculate the intensity for ${\bf r}^1$. This procedure is continued until all processors representing the upper half of the computational domain have calculated the intensities for all rays entering at the top of the cell. 
\par
The processors representing the lower half of the computational domain first calculate the intensities for the rays entering at the bottom of the cell (${\bf r}^5, \ldots {\bf r}^8$). The procedure is similar to the one described above, only the start is at the processors representing the first layer from the bottom and then moving forward to the top. 
\par
After that, the processors representing the upper half of the computational domain have to continue from the bottom to the top with the rays entering at the bottom of the cell (${\bf r}^5, \ldots {\bf r}^8$) and the processors representing the lower half of the computational domain have to continue from the top to the bottom with the rays entering at the top of the cell (${\bf r}^1, \ldots {\bf r}^4$) in the way described above. 
\par
The efficiency of this algorithm depends on the relation between $P_x$ and $N_{\rm rays,x}$. The more subdomains in $x$-direction and the less ray directions entering at the top or at the bottom, the less efficient is the implementation. 
\par
The algorithm is performed componentwise, first the rays entering at the $x$-borders, then the rays entering at the $y$-borders, and finally the rays entering at the $z$-borders are considered.

%------------------------------

\subsubsection{Frequency dependent radiative transfer \label{binning}} 

The frequency dependent radiative transfer, which is important in the solar photosphere, is implemented by the opacity binning method, see \cite{Nordlund82} and \cite{Ludwig94}.
%  \cite{18b}
\par
Frequencies $\nu$ which reach optical depth $\tau_{\nu}=1$ at similar geometric depths in a one-dimensional atmosphere are grouped into the same bin. It is assumed that frequencies which reach optical depth unity at similar geometrical depth have got a similar behavior in the radiation field.
\par 
For each frequency set $\Omega_i$ a mean opacity $\bar{\chi}_i$ and a mean source function $\bar{S}_i$ are determined for which the radiative transfer equation is solved, resulting in $I_i({\bf r})$. Then the mean intensity $J_i$ for each bin is calculated. Finally, the radiative heating rate is given by the weighted sum over the bins,
\begin{equation}
Q_{\rm rad}= 4\pi \rho \sum_{i=1}^{N_{\rm bins}} \omega_i \bar{\chi}_i (J_i-\bar{S}_i).
\end{equation}
In the ANTARES code 4 or 12 bins are used. Grey simulations are carried out with the Rosseland mean opacity and the frequency-integrated Planck-function.

The procedure used for calculating the binned opacities is based on a package for computing 
opacity distribution functions by \cite{Piskunov01}. Though in principle we could also 
use this package to compute new opacities for arbitrary abundances, in the present work we
start from the precomputed tables of Kurucz, \cite{Kurucz93}, as we first want to study the possible 
improvements gained from a more refined numerical treatment of the hydrodynamics. The 
software package of Piskunov \& Kupka, \cite{Piskunov01}, also includes a code for adding continuum 
opacities to the tables of Kurucz, \cite{Kurucz93}, which only contain line opacities. Summation over
the resulting opacity distribution yields Rosseland and Planck opacities. This capability
was required since they computed opacity distribution functions for chemically peculiar stars
for which the precomputed Rosseland tables \cite{Kurucz93} could not be used.

To compute binned opacities we take a one-dimensional model atmosphere as in 
\cite{Kurucz93} --- or in principle also horizontal and temporal averages from an earlier numerical 
simulation with the same basic parameters --- to determine the Rosseland optical depth
$\tau_{\rm Ross}$ at which each bin of the opacity distribution function reaches
$\tau_{\nu}(\tau_{\rm Ross}) = 1$. Following this analysis each bin is assigned to its
corresponding optical depth bin. On this basis we calculate partial sums of frequencies to which these 
bins correspond as well as of the Planck function and its temperature derivative,
in addition to Rosseland and Planck means. By performing a summation of these quantities
over all frequencies of interest (from about 9~nm to 160~$\mu$m) we obtain the frequency
weight, Rosseland and Planck means as well the partially summed Planck function and
its derivative with temperature for each optical depth bin. Rosseland and Planck opacities
for the entire frequency range are computed on the fly and compared with a computation
based on integrating the optical depth bins in the same manner (harmonic mean weighted
with the temperature derivative of the Planck function for the former and an arithmetic mean
weighted with the Planck function but excluding contributions from scattering for the latter).
For the tables used in our simulations these averages agree reasonably well. For the case
of Rosseland opacities we have also successfully compared our results to the original tables of Kurucz, \cite{Kurucz93}, as was done already in \cite{Piskunov01} for the case of A and
B type stars. Overall, our procedure is very similar to that one discussed in detail in
\cite{Voegler05}. 

For the opacities eventually assigned to each optical depth bin $i$ we go one step
beyond the procedure just described and follow \cite{Ludwig94} by performing
a weighted average of Planck and Rosseland opacity as a function of depth. Thus,
\begin{equation}  \label{Ludwig_average}
   \overline{\chi}_i  = {\rm e}^{-\tau_{{\rm Ross},i}} \, \overline{\chi}_{{\rm Planck},i} +
      \left(1 - {\rm e}^{-\tau_{{\rm Ross},i}}\right)\, \overline{\chi}_{{\rm Ross},i} 
\end{equation}
where $\tau_{\rm Ross}$ is estimated from a simplified hydrostatic equilibrium relation:
$\tau_{{\rm Ross},i} \approx \overline{\chi}_{{\rm Ross},i} p / g$. This puts the transition
region between Rosseland opacities (high optical depth) and Planck opacities around
$\tau_{\rm Ross} \approx 1/{\rm e}$. This setting has been used in all the calculations shown 
here, but our binning code also allows changing the location of this transition. In principle,
this parameter can be optimized together with the number of bins and their inner
boundaries. 

In our simulations for grey radiative transfer we use the Rosseland mean computed
from these tables. For the case of simulations in two spatial dimensions and non-grey
radiative transfer we mostly use 12 bins delimited by the set of points $\log \tau_{\rm Ross} = 
+0.25, 0.0, -0.25$, $-0.5, -1.0, -1.5, -2.0, -2.5, -3.0$, $-4.0, -5.0$, while for simulations in three
spatial dimensions we mostly use four bins delimited by $\log \tau_{\rm Ross} = -0.5, -1.5,
-2.5$, as in \cite{Ludwig02} for the case of cool stars. 

%---------------------------------------- RUNGE-KUTTA ----------------------------------------
\subsection{Temporal integration \label{section-rk}}

To advance the numerical solution in time either a second or a third order accurate Runge-Kutta scheme, \citet{Shu88}, is used. These time integration schemes are specifically designed for use in conjunction with ENO spatial discretisations with the aim of not introducing unwanted additional spatial variation.

%--------------------------------------------------------------------------------------------
%---------------------------------------- 2D RESULTS ----------------------------------------
%--------------------------------------------------------------------------------------------
\section{Comparison of the numerical methods}

In order to get insight on the quality of various numerical methods, we calculate two-dimensional models of solar granulation, physically identical, with different numerical schemes. Due to the dimensional splitting approach the numerics in two and three dimensions is as similar at possible at all. Hence it seems reasonable that relative advantages and disadvantages of various numerical schemes carry over from 2D, where such experiments are cheaper, to 3D. -- In addition, we compare two 3D models.

%---------------------------------------- GOAL ----------------------------------------
\subsection{Simulations}

Here we discuss the influence of numerical methods for the hyperbolic part (WENO-5, ENO-3 and CNO-4) including artificial diffusivities (AD) with WENO-5 and ENO-3 methods without AD to 
investigate the influence of the order of the methods used for the hyperbolic part of the entire
problem. We also consider the influence of AD per se (WENO-5 with AD and WENO-5 with
a minimal Prandtl number). The Prandtl number ${\rm Pr}$ is an important similarity parameter
for fluid flow:
\begin{equation}
{\rm Pr} = \frac{c_p \mu}{\kappa} \label{prandtl},
\end{equation}
$\mu$ is the dynamic viscosity, $\kappa$ is the thermal (radiative) conductivity and $c_p$ is the 
specific heat at constant pressure. The right hand side of (\ref{prandtl}) can be rewritten as
$\frac{\mu/\rho}{\kappa/(\rho c_p)}$. Hence, the Prandtl number is the ratio of the kinematic
viscosity to the thermal (in our case actually radiative) diffusivity. 
\par
\begin{figure}[ht]
\begin{center}
\includegraphics[width=55mm, angle=270]{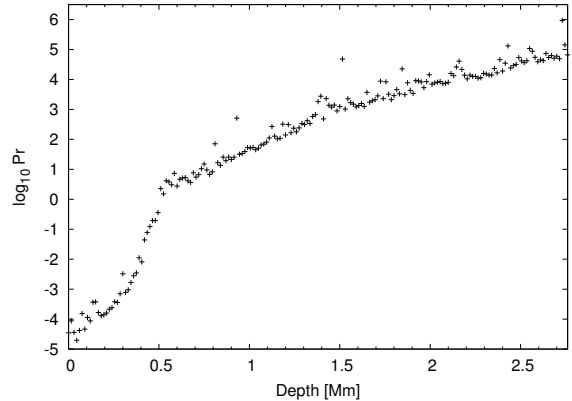}
\caption{Estimate for $\log_{10}$ of the formal Prandtl number defined from comparing
the artificial diffusivities with the viscous stress tensor.}
\label{prandtl-est}
\end{center}
\end{figure}

Through the comparison of the entries $\underline{\tau}_{kl}$ of the viscous tensor,
\begin{equation}
\label{eq-form-Pr0}
{\rm Pr} \frac{\kappa}{c_p} \left(
\partial_{x_l}u_k+\partial_{x_k}u_l-\frac{2}{3} \delta_{kl} (\nabla \cdot {\bf u})
\right)
\end{equation}
 and the corresponding component of the artificial viscosity,  
\begin{equation}
\label{eq-form-Pr1}
\frac{1}{2} \rho \left(\nu_l(u_k)\partial_{x_l}u_k + \nu_k(u_l)\partial_{x_k}u_l
\right)
\end{equation}
one can estimate a formal ``Prandtl number'' ${\rm Pr}$ which in fact is rather
a Peclet number defined for the scale of the artificial diffusivities: 
${\rm Pr_{ad}} := {\rm Pe_{ad}} = \nu_{\rm ad} / (\kappa / (c_p \rho ))$.
Figure~\ref{prandtl-est} shows this estimate for the initial configuration from which we begin 
our  further comparisons of the two-dimensional numerical simulations. At the top this
formal Prandtl number is a few times $10^{-5}$, increases rapidly and attains a value of $10^{5}$ at the bottom. 
This demonstrates that near the top a negligible amount of heat is transported at the grid
scale and confirms that for the surface layers the physically important contributions to
energy transfer are resolved on the computational grid. However, since our simulations
have to neglect several decades of scales down to those of viscous dissipation, once
we consider layers deep inside the convection zone, even at the length scale of the 
computational grid convection is found to be much more efficient in transporting heat than 
radiation and we expect a huge increase of the formal Prandtl number as is indeed found in
Fig.~\ref{prandtl-est}. From Eq.~(\ref{eq-form-Pr0})  we see that for constant ${\rm Pr}$ the 
contribution by the viscous stress tensor $\underline{\tau}_{kl}$ declines with increasing depth, 
since the fraction ${\kappa}/{c_p}$ declines rapidly and the fields of the velocity gradients do
not vary significantly inside the computational domain. On the other hand, the entries in the 
artificial diffusion tensor, Eq.~(\ref{eq-form-Pr1}), increase, since the value in the brackets, which shows the same 
behaviour inside the computational domain, is multiplied by the mass density. 
Since the Prandtl number in the solar photosphere is about $10^{-10}$, see subsection \ref{dovf}, the artificial diffusivities serve only stabilize the code and do by no means describe the physical viscous 
behavior. Therefore, a comparison of simulations with and without artificial diffusivities is 
essential. As the numerical viscosity of the WENO-5 and ENO-3 schemes themselves is 
difficult to compute and it is, moreover, not clear whether it can at all  be reasonably described by viscosity terms of the usual form, we resort to a comparison of simulation results and their dependency from
the order of the numerical method used and the presence or absence of AD in the simulation.

%---------------------------------------- SIMULATION SETUP ----------------------------------------
\subsection{Simulation setup}

The vertical and horizontal extents of the computational domain are 2763~km $\times$ 
11172~km. This domain is provided with 182 $\times$ 485 cells which yields a mesh size of 15.2~km in the vertical and 23.0~km in the horizontal direction. This resolution is common for simulations of the surface layers of the Sun. Approximately 80 minutes of solar granulation are simulated.

%---------------------------------------- RESULTS ----------------------------------------
\subsection{Results}

\begin{figure}[ht]
\begin{center}
\includegraphics[width=55mm, angle=270]{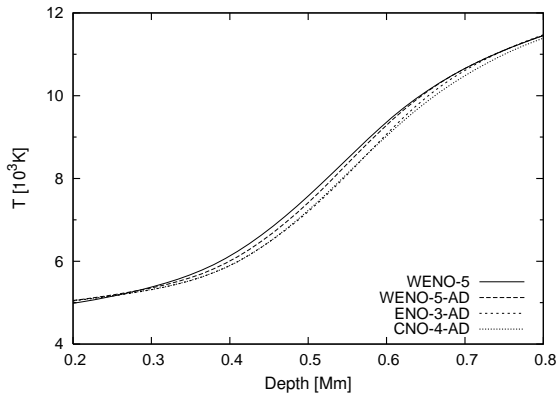}
\caption{Vertical temperature profile. Note the shift of up to about 30~km, or roughly
two grid cells, within the region where the temperature gradient is steepest.}
\label{s2f-t}
\end{center}
\end{figure}

The temporal means of the horizontally averaged temperature show small differences 
which originate from the advection schemes used for each simulation (Fig.~\ref{s2f-t} 
displays the region containing the superadiabatic layer). Among the simulations with 
artificial diffusivities, the ENO-based schemes produce steeper gradients, which 
agrees with the observations made for the test problem in Sect.~\ref{testproblem}. This 
effect is obvious when comparing the solutions for ENO-3-AD and CNO-4-AD. The 
simulation without artificial diffusivities yields a different temperature profile: a steeper 
gradient in the upper part (at the solar surface) and a flatter gradient in the lower part
(underneath it).
\par
\begin{figure}[ht]
\begin{center}
\includegraphics[width=55mm, angle=270]{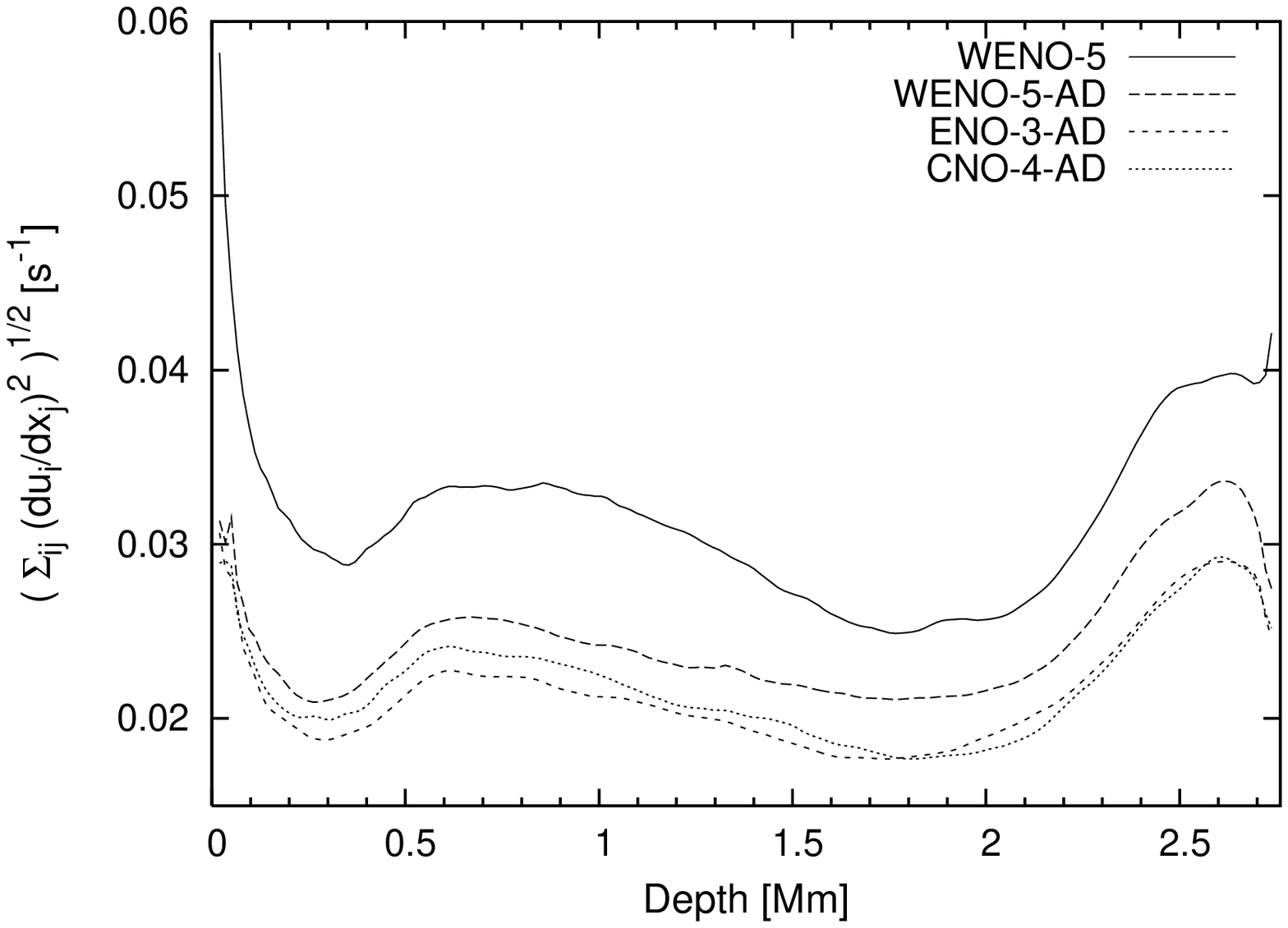}
\caption{The quantity $\left( \overline{\sum_{i,j}{\left(\partial_{x_j}u_i\right)^2}} \right)^{\frac{1}{2}}$
 is a measure for both the turbulent behaviour and
the presence of shock fronts in the system, since it is sensitive to steep gradients
caused by shear and by strong compression or expansion flows.}
\label{s2f-duidxj}
\end{center}
\end{figure}

\begin{figure}[ht]
\begin{center}
\includegraphics[width=55mm, angle=270]{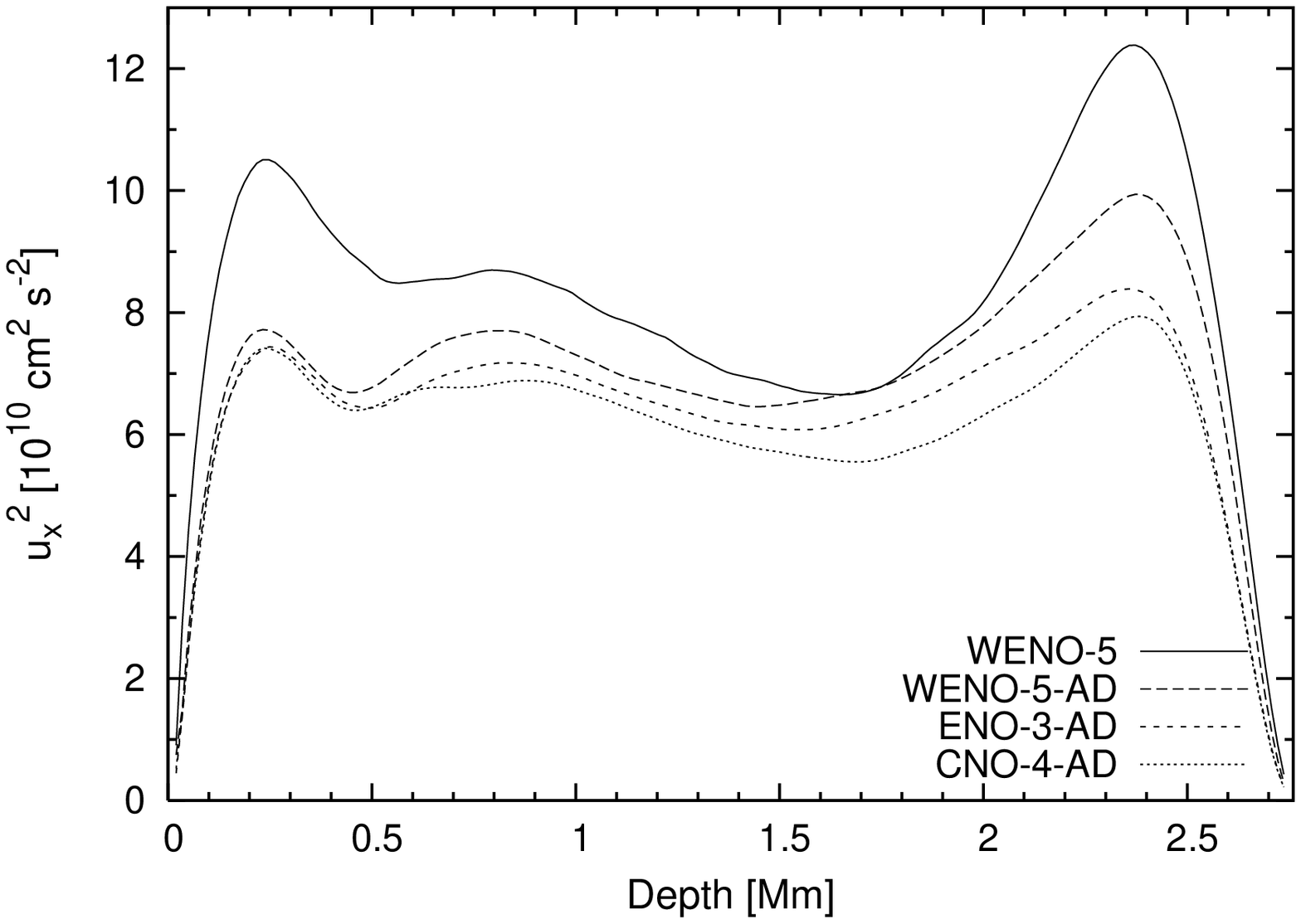}
\caption{The quantity $ {\bf u}_x^2 \equiv u^2$, the squared vertical velocity, is found
to be slightly higher in the simulation without artificial diffusivity. Note that there is no simple connection to kinetic energy flux as the latter depends on flow assymetry. }
\label{s2f-vxq}
\end{center}
\end{figure}

The temporal means of the horizontally averaged velocity components and their
spatial derivatives (see Fig.~\ref{s2f-duidxj} and~\ref{s2f-vxq} for two examples) are
similar for all three methods using artificial diffusivities. The shapes of these
curves are all similar, though their magnitudes are slightly different. Although the simulation without artificial diffusivities also yields similar averaged velocity profiles,
the numerical values obtained from this method
are clearly higher. It both yields greater shear stresses and velocity divergences
(Fig.~\ref{s2f-duidxj}) as well as higher vertical velocities (Fig.~\ref{s2f-vxq}).
\par

At a depth of 600~km the granulation simulation is comparable with solar observations
for determining the mean size of the granules. The mean distance $d$ between
granular centers, the cell size, was measured to be 1.94'' in \cite{Roudier86},  
while in \cite{Bray77} a value of $d = 1.76''$ is reported, which corresponds to 1400~km and 1270~km, respectively. Although the horizontally averaged quantities are similar for 
the simulations without artificial diffusivities, there is a difference in the mean number of 
downflows $N_{\rm DF}$ and the resulting mean cell size $d$ at a depth of 600~km
(see Table~\ref{granule_sizes}).
For the chosen resolution none of the three methods using artificial diffusivities yields
an appropriate number of downflows. Only for the simulation with a plain fourth-order discretization of the viscous stress tensor (in addition to its intrinsic, small numerical
diffusivity) one obtains an acceptable mean number of downflows and a corresponding
granule size. In evaluating this result one should keep the 2D nature of the simulations
in mind, which may not, in such detail, be comparable to observations. However, our 3D 
simulations result also in smaller granules for even finer grid spacing. Hence, these
results may also be taken in favor of WENO-5 without AD.

\begin{table}[ht]
\begin{center}
\caption{Number of downflows $N_{\rm DF}$ and granule size for the 2D simulations
using different advection schemes.} \label{granule_sizes}
\begin{tabular}{lcc}
& $N_{\rm DF}$ & $d$ [km]  \\
\hline
WENO-5    & 8.1 & 1380  \\
WENO-5-AD & 7.1 & 1580  \\
ENO-3-AD  & 6.2 & 1810  \\
CNO-4-AD  & 6.8 & 1650  \\
\hline \\
\end{tabular}
\end{center}
\end{table}

\begin{figure}[ht]
\begin{center}
\subfigure[WENO-5]{

\includegraphics[width=80mm, height=20mm]{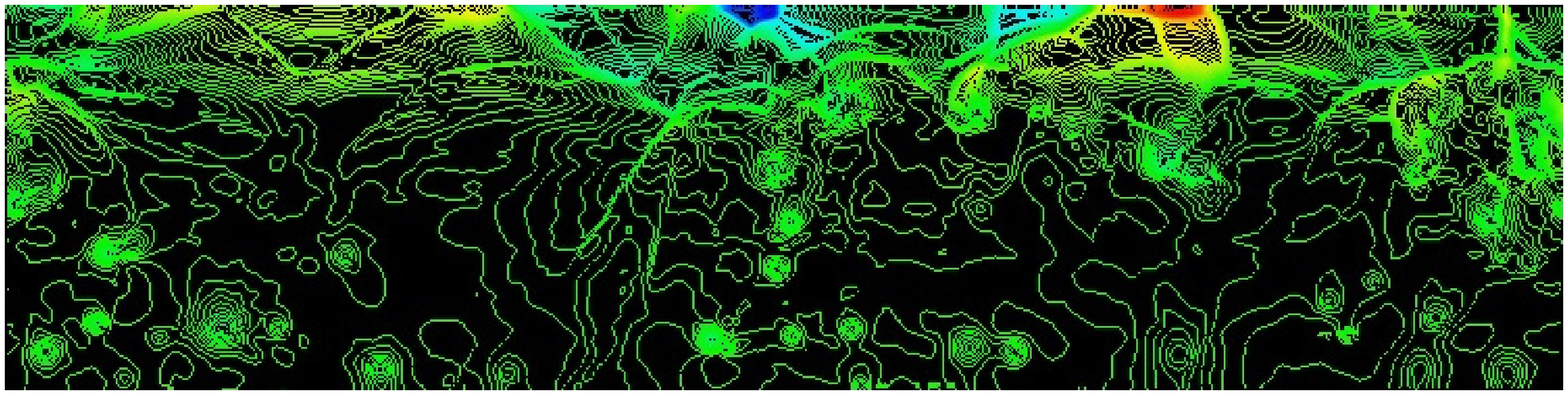}
}
\subfigure[WENO-5-AD]{
\includegraphics[width=80mm,height=20mm]{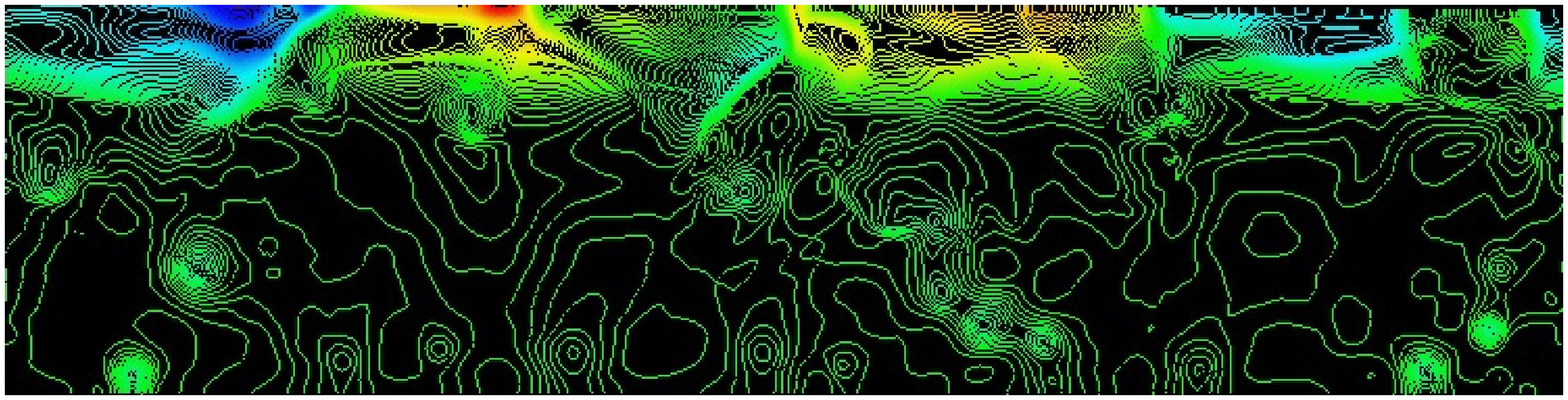}
}
\subfigure[ENO-3]{
\includegraphics[width=80mm, height=20mm]{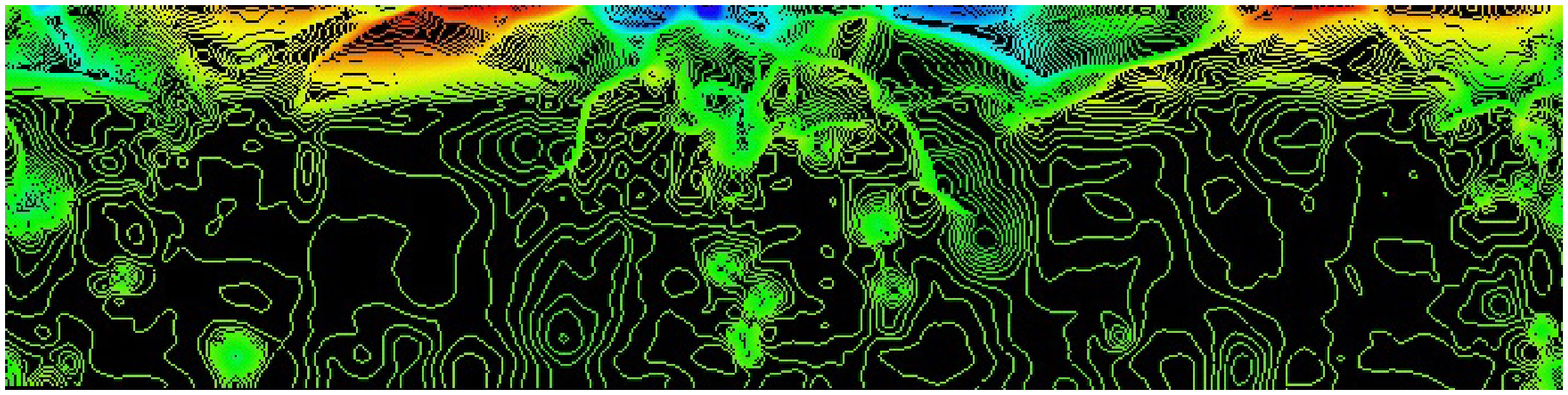}
}
\subfigure[ENO-3-AD]{
\includegraphics[width=80mm, height=20mm]{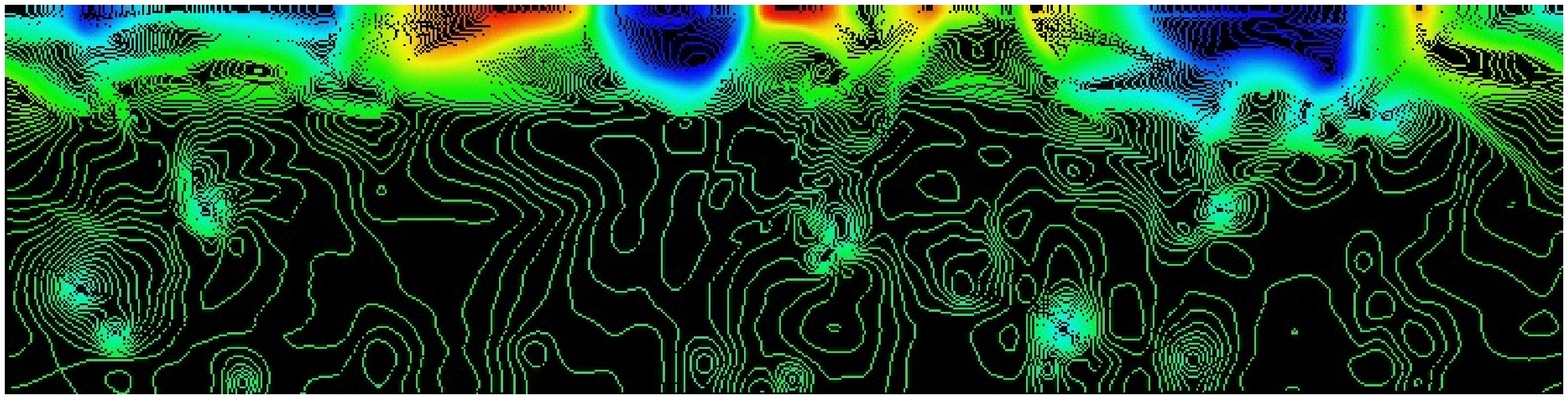}
}
\subfigure[CNO-4-AD]{
\includegraphics[width=80mm, height=20mm]{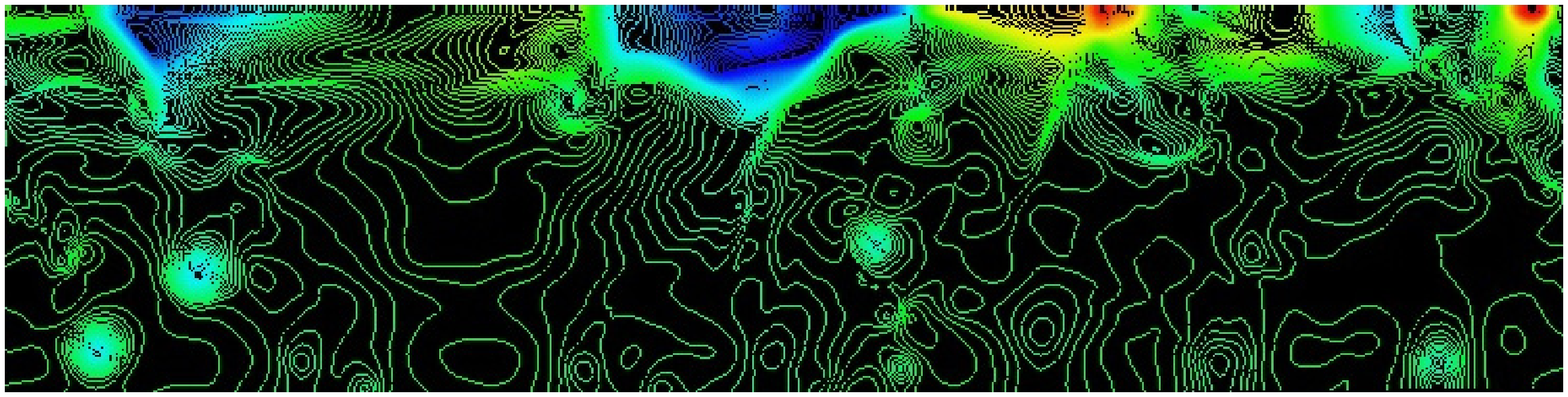}
}
\caption{Isolines of perturbations in the pressure distribution (logarithm of pressure
with logarithm of its horizontal average subtracted) after 25 minutes of evolution from
a thermally relaxed model. The isolines crowd near shocks and acoustic pulses.}
\label{s2f-p}
\end{center}
\end{figure}

In addition to one-dimensional averages and the size of the granules we can also
consider the pressure distribution (logarithm of pressure with logarithm of its horizontal average subtracted, see Fig.~\ref{s2f-p}). The simulations with this typical resolution
for solar granulation simulations of 15.2~km vertically and 23~km horizontally 
can be compared with the high-resolution model described in \cite{Muthsam07}.
The acoustic pulses which are ubiquitous in the high-resolution model are present in the simulation without artificial diffusivities. In the simulations with artificial diffusivities these acoustic pulses are not that sharp and numerous and not found so deeply down. Furthermore, the simulations without artificial diffusivities show more small scale features (e.g. at the upper right corner) and more whirls which we have also observed in our 
higher resolution runs.
\par

%---------------------------------------- DISCUSSION ----------------------------------------
\subsection{Discussion}
\label{sec:numerics-discuss}
For simulations with higher resolution and artificial diffusivities the values for the 
horizonal averages increase and the mean size of the granules decreases 
such that the differences caused by various implementations of the viscosity
(or, from a different point of view, of the modeling of the unresolved scales) vanish. A 2D high-resolution model of solar granulation with artificial diffusivities in \cite{Muthsam07} shows agreement on an quantitative level: the size of the granules is appropriate.

For 3D simulations we mostly use the fifth-order WENO scheme
with Marquina's flux splitting and entropy fix. This method is stable for both 
implementations of the viscosity (numerical viscosity plus Navier-Stokes type viscosity only, or
with artificial diffusivities included) and  the mean size of the 
granules is nearest to the observed values for the 2D simulations already at a typical 
resolution for solar granulation calculations of about 20~km. Moreover, if we consider a horizontal 
profile, where the density shows one smooth peak and at this peak the vertical velocity changes sign, we find that the temporal development of the density calculated with Marquina's flux splitting and the entropy fix remains smooth, whereas the solution calculated with the standard ENO- or WENO-method generates oscillations in the
smooth peak of the density. These observations also correspond to the theoretical considerations described in \cite{Fedkiw98}. Thus, all subsequent simulations use WENO-5 and include Marquina's flux splitting 
and entropy fix.
\par
\begin{figure}[t]
\begin{center}
\subfigure[WENO-5]{
\includegraphics[width=65mm, height=55mm]{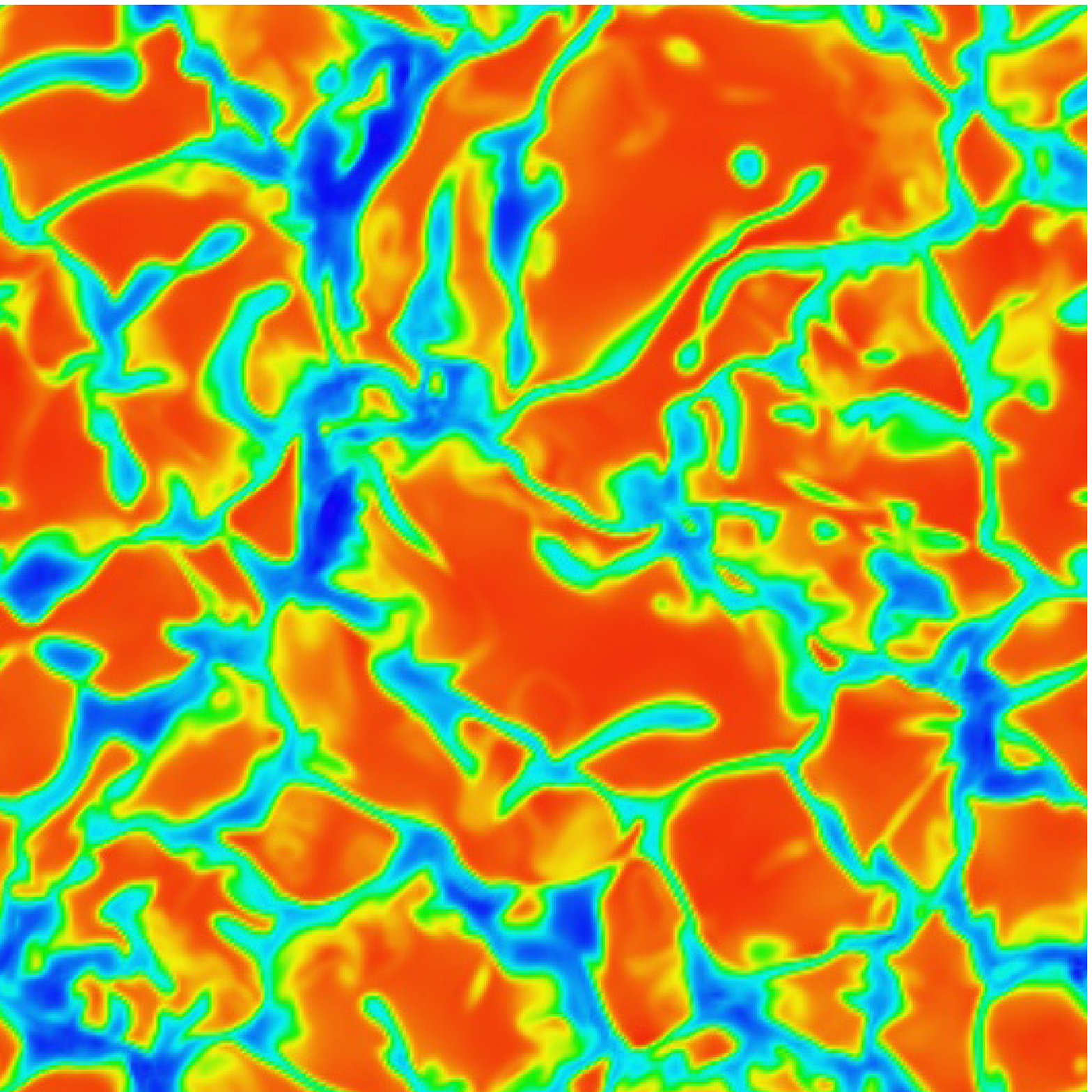}
}
\subfigure[WENO-5-AD]{
\includegraphics[width=65mm, height=55mm]{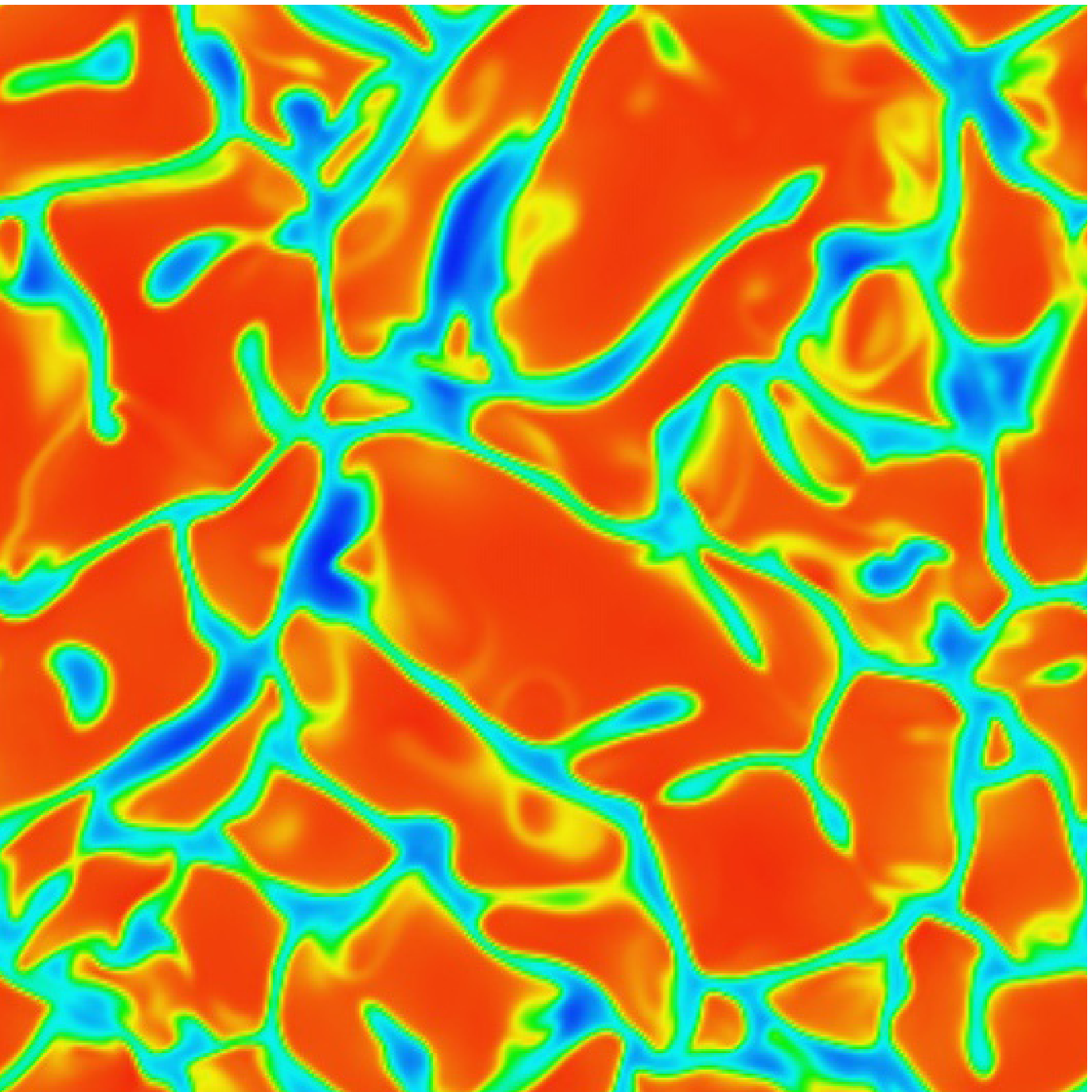}
}
\caption{Horizontal slice of the temperature distribution at a depth of 600~km seven minutes after the simulation began with the same initial model calculated with (a) WENO-5 and (b) WENO-5-AD. The temperature range is 5500~K (blue) to 11500~K (red).}
\label{s3b-t}
\end{center}
\end{figure}

If we repeat some of our numerical experiments with these settings in 3D, we again
find that a simulation without artificial diffusivities using the same mesh size yields a markedly 
higher resolution than its counterpart using artificial diffusivities. This is evident from
the horizontal slices at a depth of 600~km for 3D simulations shown in 
Fig.~\ref{s3b-t}, where more small scale features become visible. For those two
simulations the computational domain comprises 2763~km $\times$ 11172~km
$\times$ 11172~km and is provided with 182 $\times$ 285 $\times$ 285 cells which 
yields a cell size of 15.2~km $\times$ 39.2~km $\times$ 39.2~km.
Given the veracity of the methods without artificial diffusivity as discussed in the 2D case above we feel justified to trust the results in 3D to a high degree as well. 
\par
If we apply WENO-5 with no artificial diffusivities in 3D we have to add a Navier-Stokes-type diffusion term in the momentum equation with a coefficient larger than the physical one in order to keep the code stable.  Considering an ever better resolved series of calculations, we can drastically reduce the amount of viscosity we need to add as we increase resolution.  In fact, in the refinement zone of the  3D model discussed in the next chapter we can do by adding a viscous term (of the form of usual hydrodynamic viscosity) to the momentum equation \emph{only}, the coefficient being that small that, for the corresponding Prandtl number, we have  $\mbox{Pr}<10^{-6}$  everywhere. Keeping in mind the adverse effects of added viscosity as discussed earlier in the 2D case this implies that gain in effective resolution is larger when moving to a highly resolved simulations than can be expected on basis of the grid sizes alone and that, at high resolution, WENO-5 can do  with, by any numerical standards, only a very small amount of diffusivity added. From a somewhat different standpoint, it is remarkable in itself that WENO-5 really feels a viscosity which is so small in terms of Prandtl number.
\begin{figure*}[ht]
\begin{center}
\includegraphics[width=170mm]{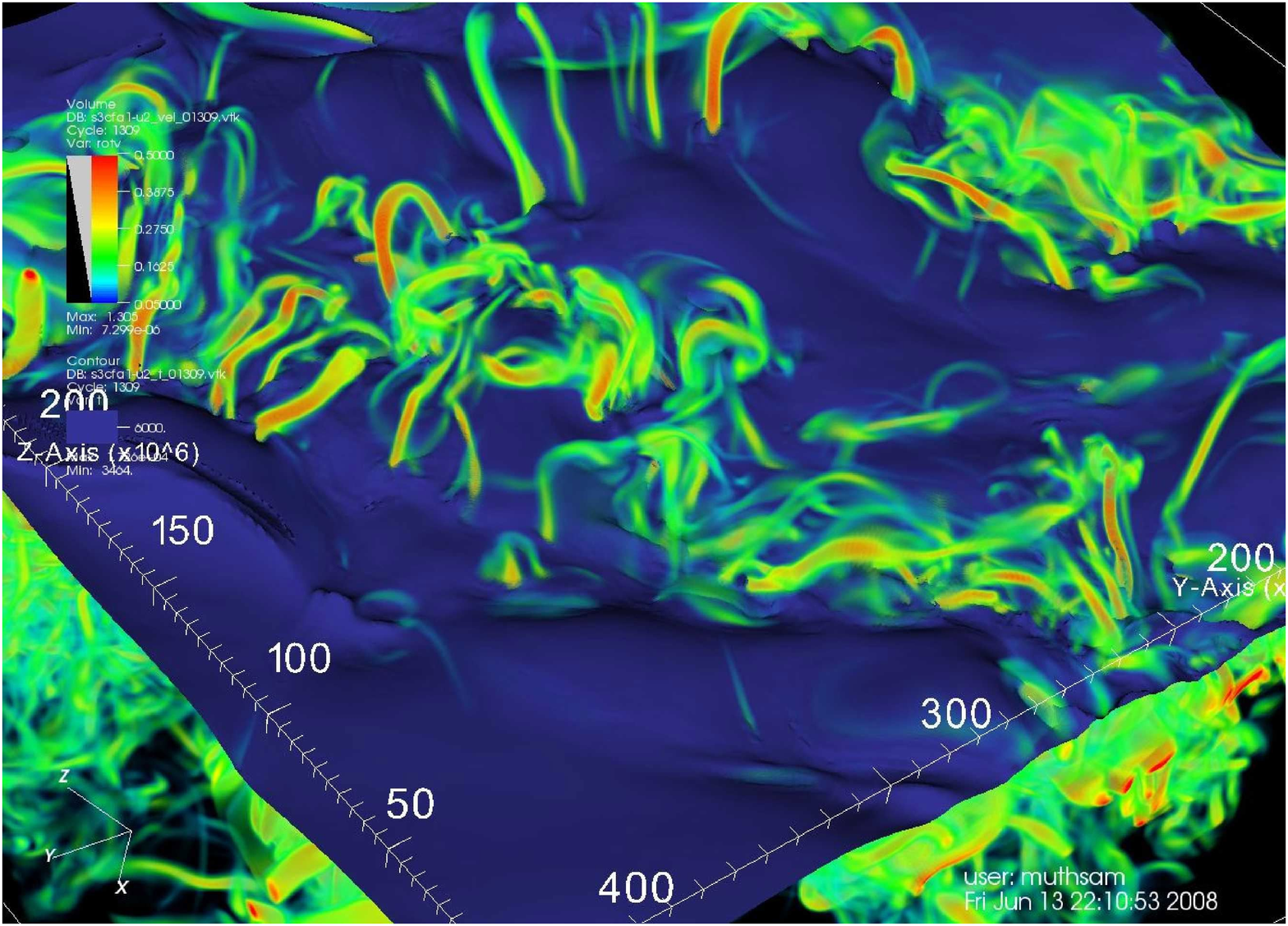}
\caption{Tornadoes in the solar atmosphere. Isosurface: $T=6000\mbox{ K}$; volume rendering: norm of vorticity. Spatial scale is in units of $10^6 \mbox{ cm}$ ($100$ corresponds to $1\mbox{ Mm}$).}
\label{tornado01}
\end{center}
\end{figure*}

\section{A three-dimensional simulation with local grid refinement}

\par
We  just have mentioned the model with the basic cube of size 2763~km $\times$ 11172~km
$\times$ 11172~km  (equipped with 182 $\times$ 285 $\times$ 285 cells,   cell size  15.2~km $\times$ 39.2~km $\times$ 39.2~km). Into that domain we have plugged a refinement zone  of size 1966~km $\times$ 4165~km $\times$ 3851~km,  provided with 259 $\times$ 425 $\times$ 393 cells. That leads to a cell size of 7.1~km $\times$ 9.8~km $\times$ 9.8~km. The high-resolution domain starts quite high in the atmosphere. We now discuss the results borne out from this high resolution simulation. Basically, our simulation follows the decay of a large exploding granule occupying the largest part of the refinement region.
\par
Quite soon, due to the higher resolution, new features become visible which were at best hinted in the lower resolution run. In fact, a considerable number of vortex tubes commenced evolving. They are generated by the downflow lanes at the boundaries of the convection cells at a depth where the downflowing material gets buoyant and is deflected sidewards and ultimately upwards.
\par
A number of these tornadoes manages to move upwards into the photosphere. They can readily be seen in Fig.~\ref{tornado01}.
The figure contains an isosurface plot of temperature, $T=6000\mbox{ K}$. The tornadoes are visualized by means of a volume rendering of the norm of vorticity. It is obvious that they crowd predominantly in the vicinity of the downflows, characterized by depressions in the $T-$isosurface, whereas the broad upflows are void of them. The tornadoes can also readily be recognized in other variables: $\nabla(p-\bar{p})$ where $\bar{p}$ denotes the horizontally averaged pressure, $\nabla(\rho-\bar{\rho)}$, etc. 
\par
The effects are large. Pressure inside the tornadoes may amount to only $\sim\frac{1}{2}$ of the ambient pressure, density may also be quite low and, in addition, temperature is lower than in the surrounding medium. The tornadoes are stabilized against collapse resulting from the higher ambient pressure by rapid rotation. 
\begin{figure*}[ht]
\begin{center}
\includegraphics[width=180mm]{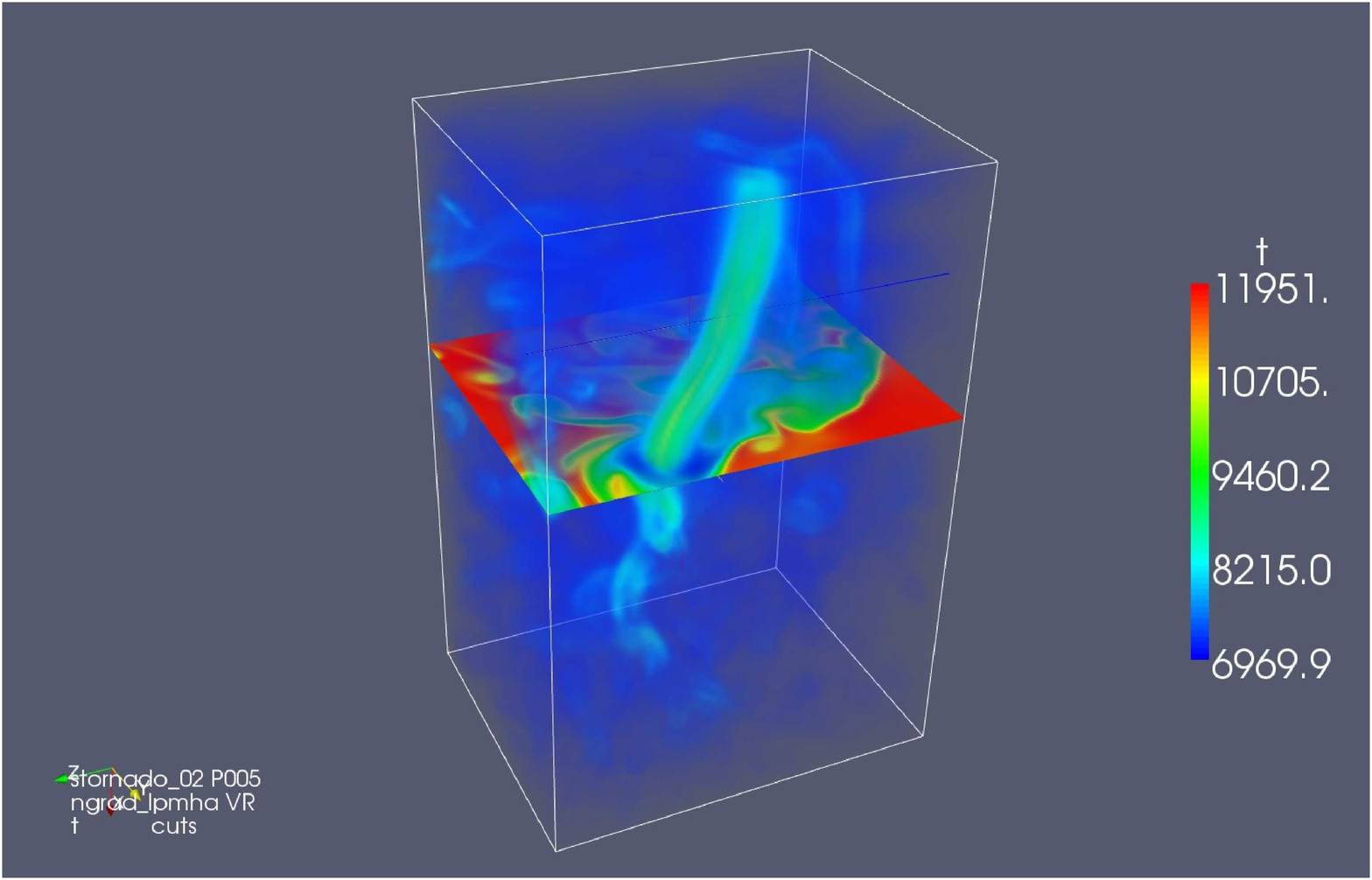}
\caption{A strong vortex tube reaching up to the high photosphere and, possibly therefore, the chromosphere. The facing side is $1\mbox{ Mm}$ wide. Volume rendering: $\nabla(p-\bar{p})$. Slice: temperature.}
\label{tornado02}
\end{center}
\end{figure*}
\par
The vortex tubes reach the surface layers by ascending predominantly at the granule boundaries where the vertical velocity is $\sim 0$ and there exists strong shear.  Typical diameters are somewhat above $100\mbox{ km}$ and smaller. Note that a tornado of that size has only about $10$ grid points to represent it. Given the strong velocity gradients it is remarkable that the tornadoes survive. It would be an interesting question to study the stability with either analytical models or numerically applying even higher resolution.
\par
These tornadoes can be considered to be the 3D analogs of the vorticity patches described, for the 2D case, by \cite{Muthsam07}. However, in moving up to the surface they behave quite differently than their 2D counterparts which stay deep below the photosphere. In fact, in 3D the tornadoes can reach up to the top of the computational domain, quite high in the photosphere, as witnessed by the strong vortex tube displayed in Fig~\ref{tornado02}. 
\par
Obviously, the tornadoes are candidates to explain the extra turbulence near downdrafts which has been diagnosed by high resolution spectroscopy of solar granulation; see e.g. \cite{Nesis93},\cite{Hanslmeier94}. As it is of interest to investigate those phenomena in ordinary granules, we have presently such work ongoing.
\par
In the paper on high resolution 2D models of solar granulation, \cite{Muthsam07}, we have observed strong acoustic pulses. Actually, they are also easily seen in Fig.~\ref{s2f-p} of the present paper which refer to the 2D case, which come from a model of much lower resolution than in the paper just mentioned. As much as resolution is being concerned their 3D counterparts should clearly show up in the present refined 3D resolution. There are, however, no such isolated and strong acoustic pulses. Rather, wavetrains can be observed (best in movies where one can actually follow them). It will require a more subtle analysis to figure out whether their energetics is comparable to that of the pulses in 2D and whether they possibly play a role for the heating of the chromosphere. 

%--------------------------------------------------------------------------------------------------
%---------------------------------------- SUMMARY and CONCLUSIONS ----------------------------------------
%-------------------------------------------------------------------------------------------------- 

\section{Summary and conclusions}
\par
We have described the ANTARES code for stellar radiation hydrodynamics whose design goals have been flexibility regarding numerical schemes and dimensions of the computation as well as microphysics (idealized or realistic). That holds also true for grid structure (straight or polar coordinates) and equations used (e.g. hydrodynamic equations in straight or polar coordinates). 

Then, we have compared various numerical schemes and demonstrated the benefits of using high-order ENO schemes. In the high-resolution runs, WENO-5 with Marquina flux splitting  stably with but a tiny amount of viscosity added in the momentum equation and without a need to add any unphysical diffusitivity in the energy equation.  This results in high information content per grid point. Grid refinement, possibly in a hierarchical fashion, allows to closer zoom in on interesting features.

In this way, we have investigated solar granulation, in particular a decaying exploding granule in the refinement region. Numerous rapidly spinning vortex tubes (diameter about $100\mbox{ km}$ or less) generated by the granule border downflows in the region where they become buoyant again manage to ascend to the photosphere near the granular downflows. Many of them are arclike, some others reach straight up to the top of the domain, i.e. high into the solar atmosphere and possibly therefore the chromosphere.

Applications of the ANTARES code to other topics in stellar physics are underway and will be reported in due time.
  
%--------------------------------------------------------------------------------------------------
%---------------------------------------- ACKNOWLEDGEMENTS ----------------------------------------
%--------------------------------------------------------------------------------------------------
\section*{Acknowledgements}
This research was supported by the Austrian Science Foundation, project P17024 and P18224. The computations were performed at the XIAN cluster of the Faculty of Mathematics, University of Vienna and at the Schr\"odinger cluster, Vienna University. We acknowledge the possibility to use computers at     MPI for Astrophysics and at RZG for calculating opacity binning tables and for part of the software development.
%--------------------------------------------------------------------------------------------
%---------------------------------------- REFERENCES ----------------------------------------
%--------------------------------------------------------------------------------------------
\newpage
%

%------------------------------------------------------------------------------------------
%---------------------------------------- APPENDIX ----------------------------------------
%------------------------------------------------------------------------------------------
%\appendix
%\section{Appendix 1}

\end{document}